\documentclass[a4paper,11pt]{article}
\usepackage{pos}
\usepackage{comment}
\usepackage{subcaption}
\usepackage{slashed}
\usepackage{dsfont}

\newcommand{\Tr}{\mathrm{Tr}}

\title{How to (Non-)Perturb a BPS Black Hole}
\author[a,b]{Alberto Castellano}
\author*[c]{Matteo Zatti}

\affiliation{$^a$Enrico Fermi Institute \& Leinweber Institute for Theoretical Physics,\\
University of Chicago, Chicago, IL 60637, USA}
\affiliation{$^b$Kavli Institute for Cosmological Physics,\\
University of Chicago, Chicago, IL
60637, USA}
\affiliation{$^c$Max-Planck-Institut f\"ur Physik (Werner-Heisenberg-Institut),\\
Boltzmannstrasse 8, 85748 Garching bei M\"unchen, Germany}

\emailAdd{acastellano@uchicago.edu, zatti@mpp.mpg.de}

\abstract{We relate the structure of non-perturbative corrections to BPS black hole observables in flat-spacetime theories with certain properties of probe charged particles in the near-horizon geometry. Concretely, we consider 4d $\mathcal{N} = 2$ supergravity with an infinite tower of F-terms and probe branes in $\text{AdS}_2\times \mathbf{S}^2$ backgrounds threaded by constant electric-magnetic fields. The higher dimensional operators we pick are computed by Type II topological string theory, and we approximate them via the constant map contribution, which is valid at large volume and can be interpreted as arising from D0-branes integrated out in M-theory on a Calabi-Yau threefold times a circle. We analyze the resulting force conditions on massive particles carrying $(q_A, p^A)$ charges, their classical trajectories, and the 1-loop effective action they produce. A simple semiclassical analysis allows us to understand qualitatively the structure of the non-perturbative corrections. The exact path integral assessment then reproduces the Gopakumar--Vafa integral of the flat-spacetime theory, now evaluated in the black hole attractor geometry. Thus, we make explicit how the physics of the fully backreacted black hole solution is controlled by the behaviour of the light D-brane states which generate the relevant set of higher derivative corrections.}

\FullConference{Proceedings of the Corfu Summer Institute 2025 "School and Workshops on Elementary Particle Physics and Gravity" (CORFU2025)\\
27 April - 28 September, 2025\\
Corfu, Greece\\}

\begin{document}

\renewcommand{\hookAfterAbstract}{
    \par\bigskip\bigskip\bigskip
    Report numbers: EFI-26-03, MPP-2026-61
}

\maketitle

\section{Introduction}\label{s:Introduction}

\noindent String theory provides a compelling framework for a consistent theory of quantum gravity. However, a complete understanding of its non-perturbative sector has remained so far elusive. In practice, our current understanding relies on a network of effective field theories (EFTs) connected by dualities, believed to emerge from a common---yet only partially understood---non-perturbative ultraviolet completion \cite{Witten:1995ex}. Some features of this UV-complete description can be encoded within the EFTs via higher derivative and higher curvature corrections \cite{Donoghue:1994dn,Donoghue:2022eay} (see \cite{vandeHeisteeg:2023dlw, Calderon-Infante:2025ldq} for a modern perspective). However, genuinely non-perturbative phenomena are typically much more difficult to access. Nonetheless, there are physical settings where these effects become apparent, due to a pronounced and perhaps surprising mixing between ultraviolet and infrared degrees of freedom. 

A striking example is provided by the seminal work of Strominger and Vafa \cite{Strominger:1996sh} where the Bekenstein--Hawking entropy of certain BPS black hole is reproduced with a microscopic counting. In order to go beyond the leading-order picture, one must incorporate quantum corrections. From the gravitational EFT perspective, perturbative contributions---organized as an infinite tower of higher derivative terms---can be systematically accounted for via Wald’s entropy formula \cite{Wald:1993nt, Iyer:1994ys}. However, as emphasized in \cite{Ooguri:2004zv, Dijkgraaf:2005bp,Dabholkar:2014ema,Iliesiu:2022kny}, even summing over all these perturbative effects may fail to fully reproduce the exact microscopic degeneracies. This mismatch indicates that genuinely non-perturbative corrections are essential for recovering a quantized black hole entropy from a macroscopic perspective.

Non-perturbative corrections can originate from several distinct physical mechanisms. Recently, the non-perturbative effects associated with the presence of charged, massive (BPS) particles in the UV theory have been further explored \cite{Castellano:2025ljk, Castellano:2025yur, Castellano:2025rvn, Castellano:2026ojb}. The picture that emerged in this series of works is that the structure of the non-perturbative corrections to BPS black hole observables in flat-spacetime theories can be related to certain properties of probe charged particles in the near-horizon geometry. The aim of this paper is to review these results and comment on related ongoing work.

\medskip

\noindent In \cite{Castellano:2025ljk}, we began by investigating the behavior of infinite towers of quantum corrections to the black hole entropy. We considered the minimal scenario which allows us to retain full computational control, namely 4d $\mathcal{N} = 2$ supergravity obtained by compactifying Type IIA string theory on a Calabi-Yau threefold. This theory admits an infinite tower of half-BPS higher derivative operators which can be related to the genus-$g$ topological free energy of the closed superstring and determined explicitly \cite{Gopakumar:1998ii,Gopakumar:1998jq}. In the large volume approximation, these corrections are dominated by the constant map contributions, which encode the quantum effects due to the integration out of D0-branes in flat spacetime. One can then build BPS black holes by solving a set of attractor equations. The corrections to the entropy of the fully backreacted geometry can be obtained by evaluating the flat-space higher derivative terms at the attractor geometry. The coupling organizing the perturbative expansion in the entropy is nothing but the ratio of two physically meaningful scales: the D0-brane/M-theory Kaluza--Klein (KK) scale and the size of the black hole.\footnote{This tells us that despite there being no instability, BPS black holes know about the KK scale.} The non-perturbative structure then emerges after performing a Borel resummation of the perturbative series, which in this case is equivalent to using the Gopakumar--Vafa (GV) line integral representation.

In \cite{Castellano:2025yur} we observed that for generic Calabi-Yau black holes the structure of the poles in the GV integral depends on the the black hole background we are considering. We identified two special instances in which non-perturbative effects are turned off. Studying the classical motion of a dyonic BPS particles in the near-horizon geometry, we characterized such cases as those in which a D0-brane probe perceives the background as purely electric or purely magnetic. In the former case, the dyonic interaction is Coulomb-like, the particle experiences a cancellation of forces and admits classical trajectories which become tangential to the boundary of the $\text{\rm AdS}_2$ throat. In all remaining cases, the classical trajectories confine the BPS particles at a finite distance, making them effectively sub-extremal, in the sense of the Weak Gravity Conjecture \cite{Arkanihamed:2006dz}. Despite the fact that these latter particles generically break supersymmetry, they also admit special trajectories on which it is restored thanks to the additional $\kappa$-symmetries of the worldline \cite{Castellano:2025rvn}. They can then induce corrections to BPS observables in the form of of worldline instantons contributions.

To go beyond the semiclassical analysis, it is necessary to evaluate precisely the 1-loop partition function associated with massive and charged particles in $\text{AdS}_2\times \mathbf{S}^2$ threaded by constant electric and magnetic fields. This was performed in \cite{Castellano:2026ojb}. We show that the supersymmetric determinant built out of a minimally coupled hypermultiplet reproduces exactly the structure of the GV integral. The role of the perturbative coupling constant is played by the ratio of the Compton wavelength of a particle with generic charges $(q_A, p^A)$ to the black hole radius. As a byproduct, we also point out that the $\mathcal{N} = 2$ hypermultiplet, which contains a Pauli-like coupling, does not seem to play the role of building block of the quantum corrections, as it was in the flat space case \cite{Dedushenko:2014nya}.

\medskip

The paper is organized as follows. In Section \ref{s:4dBHs&NonPerturbativeEffects}, we introduce the infinite tower of corrections to 4d $\mathcal{N} = 2$ supergravity that we study and discuss the non-perturbative effects induced by the relevant D-branes in generic Calabi--Yau black holes. In Section \ref{sec:semiclassicalAnalysis}, we analyze the semiclassical properties of the corresponding particle probes. This analysis is then generalized in Section \ref{s:PathIntegral}, where we evaluate exactly the one-loop determinants induced by massive and charged fields integrated out on an $\text{AdS}_2 \times \mathbf{S}^2$ background threaded by constant electric and magnetic fields. Section \ref{sec:Outlook} contains further comments and discusses future directions and ongoing work.

\smallskip

Several technical discussions are relegated to the appendices. In Appendix \ref{app:4dN2sugra}, we review our conventions for two-derivative 4d $\mathcal{N} = 2$ supergravity. In Appendices \ref{app:detailsTrajectories} and \ref{app:spectrumA2S2}, we provide additional details on the construction of classical trajectories and on the computation of the bosonic and fermionic heat kernel operators in $\text{AdS}_2$ and $\mathbf{S}^2$, respectively.

\section{4d $\mathcal{N} = 2$ Black Holes and Non-perturbative Effects}\label{s:4dBHs&NonPerturbativeEffects}

\subsection{A minimal set of higher derivative corrections to BPS black holes}\label{ss:HigherDerivativeOps}

\noindent Beyond the two-derivative theory,\footnote{See Appendix \ref{app:4dN2sugra} for a short review.} 4d $\mathcal{N} = 2$ supergravity admits an infinite tower of half-BPS higher derivative operators. These are protected F-terms which can be written as half-superspace integrals as follows \cite{Bershadsky:1993ta, Bershadsky:1993cx,Antoniadis:1993ze,Antoniadis:1995zn}
\begin{equation}\label{eq:superspacelagrangian}
	\mathcal{L}_{\rm h.d.}\, \supset\, -\frac{i}{2}\int \text{d}^4\theta\, \sum_{g\geq 1}\mathcal{F}_g (\mathcal{X}^A)\, \left(\mathcal{W}^{ij} \mathcal{W}_{ij}\right)^{g}\ +\ \text{h.c.}\, ,
\end{equation}
where $\mathcal{F}_g$ is an holomorphic and homogeneous function of degree $2-2g$, $\theta_{\alpha}$ denote the fermionic superspace coordinates (of negative chirality), $\mathcal{W}_{\mu \nu}^{ij}$ is the Weyl superfield \cite{deWit:1979dzm,Bergshoeff:1980is} and  $\mathcal{X}^A$ are the reduced chiral superfields \cite{deRoo:1980mm}. The lowest components of $\mathcal{X}^A$ and $\mathcal{W}_{\mu \nu}^{ij}$ are, respectively, the projective coordinates $X^A$ and $\frac{\epsilon^{ij}}{2} W^{-}_{\mu \nu}$, with $W^{-}_{\mu \nu}$ the anti-self-dual graviphoton field-strength whereas $\epsilon^{ij}$ is the totally antisymmetric tensor transforming under $SO(2)$. Performing the integration over the fermionic variables, one obtains contributions within \eqref{eq:superspacelagrangian} whose structure is \cite{Bergshoeff:1980is,LopesCardoso:1998tkj}
\begin{equation}\label{eq:GVterms}
	\mathcal{L}_{\rm h.d.}\, \supset\, -\frac{i}{2}\sum_{g\geq 1}\mathcal{F}_g(X^A)\, \mathcal{R}_-^2\, W_-^{2g-2}\ +\ \text{h.c.}\,.
\end{equation}
If we realize 4d $\mathcal{N} = 2$ supergravity by compactifying Type IIA string theory on a Calabi-Yau threefold $X_3$, the $\mathcal{F}_g(X^A)$ can be related to the genus-$g$ topological free energy of the closed superstring. Interestingly, as explained in \cite{Gopakumar:1998ii,Gopakumar:1998jq}, one can alternatively compute all (non-)perturbative stringy $\alpha'$-corrections exactly using the duality between physical Type IIA string theory on $X_3$ and M-theory compactified on the same threefold times a circle. This relies on the fact that the string coupling belongs to a hypermultiplet, which decouples from the vector multiplet sector at two-derivatives and thus can be freely tuned \cite{Candelas:1990pi}. A minimal set of higher derivative corrections can be then obtained by considering the constant map contribution to the topological free energy. They encode the quantum effects associated with the integration out of D0-branes. From the four-dimensional field theory perspective, one may retrieve these terms by integrating out a certain tower of spin-2, hyper- and vector multiplets in the large radius approximation of Type IIA string theory. A single hypermultiplet with central charge $Z_p = e^{K/2} \left(p^A  \mathcal{F}_{A}-q_AX^A\right)$
\cite{Bastianelli:2008cu, Dunne:2004nc} contributes to the  $\mathcal{F}_g$ via\footnote{We are using the so called D-gauge and Planck units, i.e., we set $e^{-K} = M_{\rm p}^2 = 1$.}
\begin{equation}\label{eq:generatingseries}
	\sum_{g\geq 0} \delta \mathcal{F}^{(\rm hyp)}_g (X^A)\, W_-^{2g-2} = -\frac{1}{4} \int_{i 0^+}^{i \infty}\frac{\text{d}\tau}{\tau} \frac{1}{\sin^2{\frac{\tau W_-  \bar Z_p}{2}}} e^{-\tau |Z_p|^2} \,, 
\end{equation}
with the integration being fixed along the positive imaginary axis by causality \cite{Chadha:1977my}. %The contribution of a vector multiplet is proportional to \eqref{eq:generatingseries}. 
In addition, massive gravity, vector and hypermultiplets combine in a way such that an overall factor of $\chi(X_3)$ appears \cite{Gopakumar:1998jq}. One should also note that the coupling of the superparticle to the graviphoton involves the anti-holomorphic central charge, thereby ensuring the holomorphycity of the end result \cite{Dedushenko:2014nya}. 

The deformation \eqref{eq:GVterms} of the effective action can be encoded into a \textit{generalized prepotential} via (we drop the minus sign for the graviphoton anti-self-dual part)
\begin{equation}
\label{eq:generalizedprepotential}
    F(X, W^2) = \sum_{g=0}^{\infty} F_g(X^A) W^{2g}\,,
\end{equation}
where ${F}_0(X^A) \equiv \mathcal{F}(X^A)$ is the tree-level prepotential and the $F_g$ are related to the $\mathcal{F}_g$ determined from \eqref{eq:generatingseries} via\footnote{This is the map derived in \cite{Ooguri:2004zv} relating the topological string free energy with the generalized prepotential \eqref{eq:generalizedprepotential}, adapted to our conventions. It will be crucial in what follows for determining the properties of the systems studied herein. The consistency of the emerging picture provides further evidence for the correctness of \eqref{eq:mapOSV}.} 
\begin{equation}\label{eq:mapOSV}
    F_g(X^A) = (-1)^g\, 2^{-6g} \mathcal{F}_{g} (X^A) \,.
\end{equation}
Similarly, we can define a \textit{generalized K\"ahler potential} 
\begin{equation}
      e^{-\mathscr{K}} = i \bar{X}^A F_A(X, W^2) - i X^A \bar{F}_A (\bar{X}, \bar{W}^2)\,. \label{eq:generalizedkahlerpot}
\end{equation}

\medskip

An interesting class of geometrical objects that one can construct within these theories are supersymmetric black holes.\footnote{A comprehensive treatment of this type of solutions can be found e.g., in \cite{Mohaupt:2000mj,Moore:2004fg}.} In this context it becomes clear why one introduces the notion of generalized prepotential and K\"ahler potential. At two derivatives, BPS black holes exhibit certain universal features, such as the stabilization of the moduli fields---which couple to the electromagnetic background turned on by the black hole charges $\left(q_A, p^B \right)$---at the horizon locus, according to the so-called \emph{attractor mechanism} \cite{Ferrara:1995ih,Strominger:1996kf,Ferrara:1996dd,Ferrara:1996um}. These properties are not spoiled by the higher-curvature corrections \eqref{eq:superspacelagrangian}
\cite{Behrndt:1996jn,LopesCardoso:1998tkj,Behrndt:1998eq,LopesCardoso:1999cv,LopesCardoso:1999fsj,LopesCardoso:2000qm}. Indeed, introducing 
the rescaled (K\"ahler neutral) quantities \cite{Behrndt:1996jn,LopesCardoso:1998tkj} 
\begin{equation}
\label{eq:rescaledvars}
	Y^A = C X^A\, , \qquad \Upsilon = C^2 W^2\, , \qquad  F(Y^A, \Upsilon) = C^2 F(X^A, W^2) \,,
\end{equation}
where $C^2= e^{\mathscr{K}} \Bar{\mathscr{Z}}^2$ is a compensator field with $\mathscr{Z}$ the generalized black hole central charge
\begin{equation}
     \mathscr{Z}  = e^{\mathscr{K}} ( p^A F_{A}(X, W^2) -q_A X^A)\,, \label{eq:generalizedcentralchargeBH}
\end{equation}
the attractor equations that $F$, $Y^A$ and $\Upsilon$ satisfy at the horizon are
\begin{equation} \label{eq:attractoreqs}
    p^A = 2 \, \text{Im} Y^A \,, \qquad q_A = 2 \, \text{Im} F_A \,, \qquad \Upsilon = - 64 \,. 
\end{equation}
On the other hand, the near-horizon geometry of a BPS black hole is described by \cite{Ferrara:1997yr,LopesCardoso:1998tkj}
\begin{equation} \label{eq:nearHorizonmetric}
    ds^2= -e^{2 U(y)}dt^2+ e^{-2U(y)} \left( dy^2 + y^2 d\Omega_2^2\right)\, , \qquad \text{with}\ \ e^{-2U(y)} =\frac{|\mathscr{Z}|^2 \kappa_4^2}{8\pi y^2}\,.
\end{equation}

Accordingly, the \emph{quantum entropy formula} of a solution with near-horizon region given by  (\ref{eq:nearHorizonmetric}) takes the model-independent form \cite{LopesCardoso:1998tkj} 
\begin{equation}\label{eq:entropy}
    \mathcal{S}_{\rm BH} = \pi \left[ |\mathscr{Z}|^2 + 4 \text{Im}\, \left( \Upsilon \partial_{\Upsilon} F(Y, \Upsilon) \right) \right]\, ,
\end{equation}
and is entirely determined by the black hole charges via \eqref{eq:attractoreqs}. The first term in \eqref{eq:entropy} coincides with the value of the horizon area divided by $4 G_N$, hence providing for the Bekenstein-Hawking contribution to the entropy, whilst the second piece captures deviations from the area law. Notice that the entropy (\ref{eq:entropy}) has been computed using Wald's prescription \cite{Wald:1993nt,Iyer:1994ys} within the restricted framework of conformal off-shell $\mathcal{N} = 2$ supergravity coupled to $n_V+1$ vector multiplets \cite{Ferrara:1977ij,deWit:1979dzm,deWit:1980lyi,deWit:1984rvr,deWit:1984wbb}. Upon doing so we might be missing some contributions present in the full Type IIA string theory (see \cite{Castellano:2025ljk, Castellano:2025yur} and references therein). In \cite{Ooguri:2004zv} a detailed analysis of the origin of the entropy above was performed. There it was shown that \eqref{eq:entropy} is the Legendre dual of the free energy $\text{Im} \, F$
\begin{equation} \label{eq:entropyLegendre}
    \mathcal{S}_{\rm BH} = 4 \pi \left[\text{Im} \, F - \frac{1}{2} q_A \text{Re}\, Y^A  \right] = 4 \pi \left[1 -  \text{Re}\, Y^A\frac{\partial}{\partial  \text{Re}\, Y^A}  \right]\, \text{Im} \, F\,, 
\end{equation}
where $q_A$ and $\text{Re}\, Y^A$ are, respectively, the electric charges and potentials of the black hole. Then, it was suggested that, in the context of the full Type IIA string theory, the free energy appearing in \eqref{eq:entropyLegendre} should correspond to a protected supersymmetric index rather than the partition function. In particular, they point out that this would explain why the formula is independent of the hypermultiplet vacuum expectation values (vevs). In what follows, we will not be concerned about whether (\ref{eq:entropy}) and \eqref{eq:entropyLegendre} are truly computing an entropy or the related supersymmetric index in the Type IIA theory, and we will just focus on their properties.

\subsection{The D0-brane effects in Calabi-Yau black holes}\label{sec:D0branesEffects}

\noindent We are interested now in the structure of D0-brane corrections to the entropy of a generic Calabi-Yau black hole. In the large volume approximation\footnote{\label{fnote:largevol}With this we mean the limit $z^a \to i\infty$ for all $a= 1, \ldots, h^{1,1}(X_3)$.}  the generalized prepotential \eqref{eq:rescaledvars} can be well-approximated by the function
\begin{equation}
\label{eq:holomorphicprepotential@largevol}
	F(Y, \Upsilon) = \frac{D_{a b c} Y^a Y^b Y^c}{Y^0} + d_a\, \frac{Y^a}{Y^0}\, \Upsilon + G(Y^0, \Upsilon)\, +\, \mathcal{O} \left(e^{2 \pi i z^a} \right)\, .
\end{equation}
The cubic term corresponds to the well-known tree-level contribution, with $D_{abc} =  -\frac16 \mathcal{K}_{abc}$ and where $\mathcal{K}_{abc}$ denote the triple intersection numbers of the threefold. The second term is the genus-1 topological string amplitude expanded around the large radius point \cite{Bershadsky:1993ta, Bershadsky:1993cx,Katz:1999xq}. Here, $d_a = -\frac{1}{24} \frac{1}{64} c_{2,a}$, where $c_{2,a}$ are the components of the second Chern class of $X_3$ tangent bundle expressed in a basis of harmonic 2-forms of $ H^{1,1}(X_3, \mathbb{Z})$. This contribution can be easily understood as coming from the dimensional reduction of an analogous four-derivative, curvature squared operator in 5d $\mathcal{N}=1$ supergravity \cite{Grimm:2017okk}. The last term encodes perturbative D0-brane effects. Indeed, for $g\ge2$, the leading contribution in the large volume patch corresponds to constant maps from the worldsheet to the Calabi--Yau threefold \cite{Bershadsky:1993cx}. Specifying \eqref{eq:generatingseries} for a tower of D0 bound states, taking into account the map \eqref{eq:mapOSV} and performing a genus expansion of $G(Y^0, \Upsilon)$, one obtains the asymptotic series
\begin{equation}\label{eq:Gfn}
	G(Y^0, \Upsilon) = -\frac{i}{2 (2\pi)^3}\, \chi_E(X_3)\, (Y^0)^2 \sum_{g=0, 2, 3, \ldots} c^3_{g-1}\, \alpha^{2g} + \ldots\, ,
\end{equation}
where we defined
\begin{equation}\label{eq:Gfactors}
	c^3_{g-1} = (-1)^{g-1} 2 (2g-1) \frac{\zeta(2g) \zeta(3-2g)}{(2\pi)^{2g}}\, , \qquad \alpha^2 = -\frac{1}{64}\, \frac{\Upsilon}{(Y^0)^2}\, .
\end{equation}
The ellipsis in \eqref{eq:Gfn}, on the other hand, are meant to indicate that there would be a priori further non-analytic corrections around $\alpha=0$ \cite{Castellano:2025ljk}.

\medskip

 The general prepotential  \eqref{eq:holomorphicprepotential@largevol} does not allow us to solve the attractor equations \eqref{eq:attractoreqs} in a fully explicit way. However, this issue is already present at the tree-level approximation \cite{Shmakova:1996nz} and is not exclusive to the inclusion of higher-genus corrections. Assuming that there exists a set of $x^a$ solving $\Delta_a = D_{abc} x^b x^c$, with
\begin{equation}
    \Delta_a = - \tilde{q}_a p^0 + 3 D_{abc} p^b p^c\,, \qquad \tilde{q}_a \equiv q_a + p^0 \frac{d_a \Upsilon}{|Y^0|^2} \,,
\end{equation}
and introducing 
\begin{equation}
     \tilde{q}_0 \equiv q_0 + 2 \text{Im} \left[ \frac{1}{(Y^0)^2} d_a Y^a \Upsilon \right] - 2 \text{Im}\, G_0 \,, \qquad G_0 = \frac{\partial}{\partial Y^0} G(Y^0, \Upsilon)\,,
\end{equation}
one can show that, in the most general situation with arbitrary\footnote{Formula \eqref{eq:finalentropy} is valid as long as the chosen set of charges admits a BPS solution to the attractor equations \cite{Denef:1998sv,Johnson:1999qt,Denef:2000nb}.} $(p^A, q_A)$ charges for the black hole, the expression \eqref{eq:entropy} becomes \cite{Castellano:2025yur}
\begin{equation} \label{eq:finalentropy}
    \begin{split}
    \mathcal{S}_{\rm BH} = \; &  \frac{\pi}{3 p^0}
    \sqrt{\frac{4}{3}\left({\Delta}_a {x}^a\right)^2 - 9 \left(p^0 p^A \tilde{q}_A - 2 D_{abc}p^a p^b p^c\right)^2 }  \\[2mm] & + 4\pi \left[ \text{Im}\, G -  \text{Re}\,Y^0 \text{Im}\, G_0  - \frac{1}{\sqrt{3}} \frac{(\text{Re}\, Y^0)^2}{|Y^0|^3} {x}^a d_a \Upsilon  \right]  \,,
    \end{split}
\end{equation}
with $\Upsilon = - 64$. Despite $x^a$ being implicit, the entropy above depends only on the gauge charges and on $Y^0$. At the attractor point, $Y^0$ is just the inverse of $\alpha$, which plays the role of the (complex) coupling constant governing the perturbative higher-genus corrections to the prepotential (cfr. equation \eqref{eq:Gfn}). This implies that the entropy also admits an expansion in $\alpha$ encoding higher order contributions and each of those is solely determined by the gauge charges. Furthermore, we note that at after solving \eqref{eq:attractoreqs}, the norm and the phase of $\alpha$ can be explicitly connected with other physical quantities. In fact, the latter is related to the central charge of the black hole $Z_{\rm BH}$ and that of a single D0 brane $Z_{\rm D0}$ via
\begin{equation} \label{eq:defAlphaCC}
    \alpha^{-1} = \bar{Z}_{\rm BH} Z_{\rm D0}  \,,
\end{equation}
which implies that $|\alpha|$ is equal to the ratio of the D0 length-scale---computed from its mass---to the black hole radius $r_h$ \cite{Castellano:2025ljk} (we restore the dependence on the Planck mass) 
\begin{equation}
    |\alpha| = \frac{r_5}{r_h} \,, \qquad r_h= |Z_{\rm BH}| M_p^{-1} \,, \qquad (r_5)^{-1} = m_{\rm D0}=  |Z_{\rm D0}| M_p \,.
\end{equation}
Calabi-Yau black holes with $|\alpha| \ll 1$ are larger than the M-theory circle and can be regarded as purely four-dimensional objects. On the other hand, if we now consider the opposite situation, namely $|\alpha| \gtrsim 1$, the black hole radius becomes smaller than the M-theory circle, and thus the physics should be more adequately described within the 5d realm. The transition regime therefore corresponds to black hole solutions with $\alpha \sim 1$. Notice that in presence of D6 charge, i.e., $p^0 \ge 1$, the norm of $Y^0$ is lower bounded and the $|\alpha| \gg 1$ limit cannot be explored \cite{Castellano:2025ljk}. 

\medskip

We stress that the perturbative expansion in $|\alpha| \ll 1 $ can provide a good approximation to the exact result only if we truncate the series \cite{Shenker:1990uf,Pasquetti:2010bps}. Hence, to obtain a well-defined expression for the (universal piece of the) quantum-corrected generalized prepotential \eqref{eq:holomorphicprepotential@largevol} beyond the asymptotic expansion \eqref{eq:Gfn}, we must evaluate more carefully the one-loop calculation associated with the D0-brane tower, using its integral representation\footnote{Independently of its topological string origin, \eqref{eq:G&I(alpha)} and \eqref{eq:I(alpha)} may be obtained by applying Borel resummation to \eqref{eq:Gfn}. See Appendix A of \cite{Castellano:2025ljk} for further details.}
\cite{LopesCardoso:1999fsj, Gopakumar:1998ii}
\begin{align}\label{eq:G&I(alpha)}
	G(Y^0, \Upsilon) =\frac{i}{2 (2\pi)^3}\, \chi_E(X_3)\, (Y^0)^2\, \mathcal{I}(\alpha)\, ,
\end{align}
where 
\begin{align}\label{eq:I(alpha)}
	\mathcal{I}(\alpha)\, =\, \frac{\alpha^2}{4} \sum_{n \in \mathbb{Z}}\int_{0^+}^{\infty} \frac{\text{d}s}{s}\, \frac{1}{\sinh^2\left( \pi n \alpha  s\right)}\, e^{-4\pi^2 n^2 i s} \, .
\end{align}
This formula is well-behaved provided $\text{Re} (\alpha) \ne 0$ (see below for the special case $\text{Re} (\alpha) = 0$). On top of the perturbative corrections displayed in \eqref{eq:Gfn}, the expression \eqref{eq:I(alpha)} encodes non-perturbative effects in the complexified coupling constant $\alpha$. Indeed, the presence of poles in the complex $s$-plane implies that $\mathcal{I}(\alpha)$ can be decomposed as a line integral plus an infinite sum over residues, with the latter having a structure that is reminiscent of non-perturbative corrections in $\alpha$. In general, this decomposition can be ambiguous, since it is not clear whether the choice of contour is such that all non-perturbative physics is captured by the residues alone. However, in the case at hand the contour of integration is fixed using causality considerations (cfr. the discussion around \eqref{eq:generatingseries}) and thus we can unambiguously decompose $\mathcal{I}(\alpha) = \mathcal{I}^{(p)}(\alpha) + \mathcal{I}^{(np)}(\alpha)$. Separating positive and negative mode contributions into $\mathcal{I}_{n \ge 0}$ and $\mathcal{I}_{n < 0}$ and performing the rescaling $s \rightarrow s / n$, we find that the contour of $\mathcal{I}_{n < 0}$ lies on the negative real axis. Deforming this line integral toward the positive real axis, and ignoring the piece coming from the residues associated with its integrand, allows us to obtain the perturbative part of \eqref{eq:I(alpha)} 
\begin{align}\label{eq:I(alpha)pert}
	\mathcal{I}^{(p)}(\alpha) = \frac{\alpha^2}{4} \sum_{n \in \mathbb{Z}}\int_{0^+}^{\infty} \frac{\text{d}s}{s}\, \frac{1}{\sinh^2\left( \frac{\alpha s}{2}\right)}\, e^{-2\pi i n s}\, =\, \frac{\alpha^2}{4}\, \sum_{k=1}^{\infty} \frac{1}{k \sinh^2 \left( \frac{\alpha k}{2}\right)}\, ,
\end{align}
where in the second step we performed a Poisson resummation. The  residues of $\text{csch}^2(x)$ picked up by $\mathcal{I}_{n < 0}$ then give us the non-perturbative contribution to $\mathcal{I}(\alpha)$, which reads
\begin{equation}\label{eq:Inonpertalpha1stmethod}
	\begin{aligned}
		\mathcal{I}^{(np)}(\alpha)\, &=\, -2 \pi i \alpha \sum_{n, k=1}^{\infty} \frac{n}{k}\, e^{-\frac{4\pi^2 k n}{\alpha}} \left( 1+ \frac{\alpha}{4\pi^2 kn}\right)\\
        & =  \sum_{n\ge 1}  \left[ \frac{\alpha^2}{2 \pi i}\mathrm{Li}_2\left(e^{-\frac{4 \pi^2}{\alpha} n  }\right) - 2 \pi i \alpha n \, \mathrm{Li}_1\left(e^{-\frac{4 \pi^2}{\alpha} n  }\right)\right] \,.
	\end{aligned}
\end{equation}
Notice that the same result for $\text{Re} (\alpha) \ne 0$ can be obtained upon writing \eqref{eq:I(alpha)} as a contour integral 
\begin{equation} \label{contourIntgral}
    G(Y^0, \Upsilon) = \frac{i}{4 (2\pi)^3}\, \chi_E(X_3)\, (Y^0)^2\,\alpha^2 \oint \frac{ds}{s} \frac{1}{1 - e^{-2\pi i s}} \frac{1}{\sinh^2{\left(\frac{\pi \alpha s}{2}\right)}}\, , 
\end{equation}
%
%
%%%%%%%%%%%%%%%%%%%%%%%%%%%%%%%%%%%%%%%%%%%%%
\begin{figure}[t!]
	\begin{center}
		\includegraphics[scale=0.6]{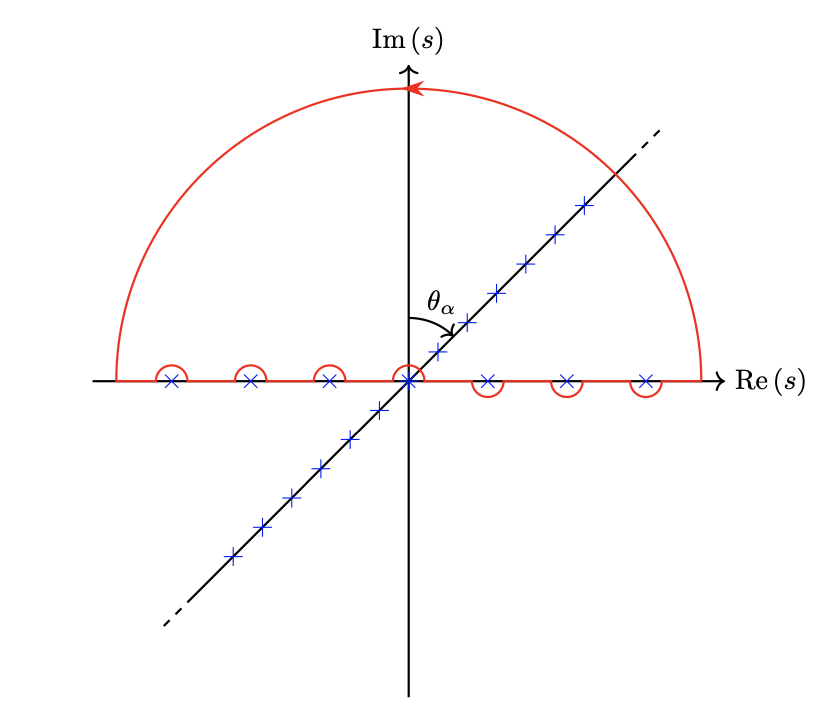}
		\caption{\small Integration contour in the complex s-plane employed to evaluate the loop integral \eqref{contourIntgral}. The non-perturbative singularities lie along $\mathbb{R}e^{i(\pi/2 - \theta_\alpha)}$, with $\theta_\alpha$ the complex phase of $\alpha$, whereas the perturbative ones fall onto the real axis. In the limit $\text{Re}\, \alpha \rightarrow 0 $, all the poles become real. %However, for the general rational values of $\alpha$, perturbative and non-perturbative singularities do not match.
        } 
		\label{fig:contourIntegralComplex}
	\end{center}
\end{figure}
%%%%%%%%%%%%%%%%%%%%%%%%%%%%%%%%%%%%%%%%%%%%%
%
where the path is specified in Figure \ref{fig:contourIntegralComplex}. In this latter formulation, we retrieve $\mathcal{I}^{(p)}(\alpha)$  and $\mathcal{I}^{(np)}(\alpha)$ directly from the residues of half of the poles located along the real and the $i\mathbb{R}\,\alpha^{-1}$ axes, respectively. In the limit $\text{Re}(\alpha) = 0$, the formula \eqref{contourIntgral} breaks down and the series of residues diverges, which can be traced back to \eqref{eq:I(alpha)pert} not being well-defined. The integration contour is now seen to cross the poles of $\cosh^2(x)$, and in order to compute \eqref{eq:I(alpha)} we carefully deform the line integrals toward the imaginary axis. Removing the $0^+$ regulator with a proper subtraction of the UV divergences, we get
\begin{equation}
\label{eq:I(alpha)selfdualmagneticintegraldef}
\begin{aligned}
    \mathcal{I}(\alpha)\, &=\, \zeta(3)\, -\frac{|\alpha|^2}{2} \sum_{n>0}\int_{0}^{\infty} \frac{\text{d}\tau}{\tau}\, e^{-4\pi^2 n^2 \tau} \left( \frac{1}{\sinh^2\left( \pi n |\alpha|  \tau\right)}- \frac{1}{(\pi n |\alpha|  \tau)^2} +  \frac13\right)\\
    &= \zeta(3)\, -\, \frac{|\alpha|^2}{2} \int_{0}^{\infty} \frac{\text{d}s}{s}\, \frac{e^{-\frac{4\pi s}{|\alpha|}}}{1-e^{-\frac{4\pi s}{|\alpha|}}} \left( \frac{1}{\sinh^2\left( s\right)} - \frac{1}{s^2}+\frac13\right)\,.
\end{aligned}
\end{equation}
Crucially, we see that despite \eqref{eq:I(alpha)selfdualmagneticintegraldef} having poles, we cannot perform any rotation that picks them, thereby reflecting the absence of non-perturbative ambiguities in the present special case.

\medskip

Let us conclude this section by commenting on various salient features exhibited by the non-perturbative corrections to the generalized prepotential, when evaluated at the attractor point. The perturbative terms \eqref{eq:Gfn} have the form of an asymptotic series with zero radius of convergence. Performing a Borel resummation of such series we obtain that the corrections remain finite for $|\alpha| \in (0, \infty)$. This means that within a one-parameter family of black hole solutions one can cross smoothly the EFT transition regime associated with a decompactification from 4d $\mathcal{N} =2 $ to 5d $\mathcal{N} = 1$ supergravity. The $|\alpha| \gg 1$ limit can be moreover explored provided the family of black holes does not carry D6-brane charge, which occurs whenever the tree-level solutions uplift to BPS black strings in five dimensions. This corresponds to the $\text{Im} (\alpha) = 0$ case, wherein the quantum-corrected black hole entropy (including just D0-brane effects) has the explicit form
\begin{equation}\label{eq:finalresummedentropynoD6}
	\begin{aligned}
		\mathcal{S}_{\text{BH}}\, =\ & 2 \pi \sqrt{\frac{1}{6} |\hat{q}_0| \left( \mathcal{K}_{abc} p^a p^b p^c + c_{2,a}\, p^a \right)} \left( 1 -  \frac{\chi_E(X_3)\,Y^0\, \alpha^2}{(2\pi)^3 |\hat{q}_0|}\, \sum_{n=1}^{\infty} \frac{n^2\, e^{-\alpha n}}{1-e^{-\alpha n}}\right)^{-1/2}\\
		&-\frac{\chi_E(X_3)}{4\pi^2} (Y^0)^2 \alpha^2 \left( \sum_{n=1}^{\infty} n \log\left(1 - e^{-\alpha n}\right) - (Y^0)^{-1} \sum_{n=1}^{\infty} \frac{n^2 e^{-\alpha n}}{1 - e^{-\alpha n}} \right)\,.
	\end{aligned}
\end{equation}
It is then simple to verify that in the limit $\alpha \rightarrow \infty $ all non-local, higher-genus contributions become suppressed and one recovers exactly the (regularized) entropy of an infinitely extended string in 5d $\mathcal{N} = 1$  (see \cite{Strominger:1996sh,Maldacena:1997de, Vafa:1997gr, Harvey:1998bx} for a microscopic entropy analysis and \cite{Sen:2005wa,Kraus:2005vz,Castro:2007sd,deAntonioMartin:2012bi, Meessen:2012su,Gomez-Fayren:2023wxk} for the macroscopic one). The only surviving corrections are those associated with the genus-1 piece of the prepotential, which directly descends from the $t_8 t_8 \mathcal{R}^4$ term in 11d supergravity \cite{Antoniadis:1997eg}. Notice that this example also provides strong evidence in favour of the map \eqref{eq:mapOSV} as well as the choice of contour in \eqref{eq:generatingseries}.\footnote{Indeed, different choices might lead to interpret e.g., \eqref{eq:I(alpha)selfdualmagneticintegraldef} as the relevant corrections for a black string. However in such case there is no dilution of the higher-genus corrections for $|\alpha| \gg 1$, which is crucial for the microscopic matching.} 

\medskip

Finally, we note that, by focusing our attention on the free energy $\text{Im}\, F$, we can already identify two configurations whose residues structure is special. On the one hand, we have the setups with $\text{Re}(\alpha) = 0$. As already discussed around equation \eqref{eq:I(alpha)selfdualmagneticintegraldef}, poles arise in the complex $s$-plane; however there exists no contour deformation that allow us to pick them up. On the other hand, we have those configurations with $\text{Im}(\alpha) = 0$. In this case, the generalized prepotential receives a purely real correction from the non-perturbative residues, which are hence invisible to the free energy $\text{Im}\, F$. A natural question then is whether we can give a physical interpretation of this observation. Comparing with equation \eqref{eq:defAlphaCC} we notice that the second case is the one in which the D0 branes have their central charge aligned with that of the black hole background. Instead, in the first scenario the two central charges exhibit a relative phase of $\pm \pi/2$. This suggests a clear link between the structure of the corrections and the properties of the particles that produced them. In the next sections, we will explore this connection further from the semiclassical perspective.

\subsection{Semiclassical analysis}\label{sec:semiclassicalAnalysis}

\noindent To understand why there appear to be cases where non-perturbative corrections to the quantum entropy formula \eqref{eq:entropy} are absent for certain BPS states in the theory, it is necessary to fully grasp how the black hole background is perceived by the latter. Only then we may hope to find what distinguishes those systems from the most general situation, where these effects seem to be generically present. To accomplish this, a good starting point would be to study the semiclassical dynamics of probe BPS particles in the two-derivative black hole geometry. This is the aim of this section. Let us stress that even though a more robust analysis would require to perform the exact quantum path integral associated with these massive fields in AdS$_2\times \mathbf{S}^2$ (cfr. Section \ref{s:PathIntegral}), many of the most relevant features can be understood already from a semiclassical point of view.

\medskip

A probe (massive) particle with charges $(q_A, p^A)$ 
propagating in the near-horizon geometry of a BPS black hole couples to a single effective constant gauge field aligned with the graviphoton. Using Poincaré coordinates and Planck units, the geometry \eqref{eq:nearHorizonmetric} becomes
\begin{equation}\label{eq:Poincaremetric}
 ds^2=\frac{R^2}{\rho^{2}}\left( -dt^2+d\rho^2\right)+ R^2d\Omega_2^2\, ,\qquad \text{with}\quad R^2\,=|Z_{\rm BH}|^2\, .
\end{equation}
The bosonic part of the 1d worldline action of such a probe particle \cite{Billo:1999ip,Simons:2004nm} then reads\footnote{We have changed slightly the convention compared to that of \cite{Castellano:2025yur,Castellano:2025rvn} when writing down the action \eqref{eq:BPSwordlineactionAdS2xS2}. In particular, the relative factor of 2 between the kinetic and gauge terms in $S_{wl}$ now appears within the latter.} 
\begin{equation}\label{eq:BPSwordlineactionAdS2xS2}
 S_{wl} = -m  \int_\gamma d\sigma R \,\sqrt{\rho^{-2}\left(\dot{t}^2-\dot{\rho}^2\right) -\dot{\theta}^2-\sin^2\theta \dot{\phi}^2} + \frac12\int_\Sigma p^AG_A-q_A F^A\, ,
\end{equation}
where $\dot{x}^{\mu} := dx^\mu/d\sigma$. Here, $\sigma$ is any convenient parameter along the spacetime trajectory, which we denote by $\gamma$, whereas $\Sigma$ corresponds to any 2d surface that ends on the worldline. The $G_A$ are the magnetic duals of the field strengths $F^A$, which are related via the linear constraint $G_A^-=\bar{\mathcal{N}}_{AB}F^{B,\, -}$, with $\mathcal{N}_{AB}$ given by \eqref{eq:Nab}. In the presence of a black hole with charges $(q_A{}',p^A{}')$, their profile can be obtained from the relations
\begin{equation}\label{eq:BHcharges}
 p^A{}'=\frac{1}{4\pi}\int_{\mathbf{S}^2} F^A\, ,\qquad  q_A'=\frac{1}{4\pi}\int_{\mathbf{S}^2} G_A\, .
\end{equation}
%
\begin{comment}
The $U(1)$ field strengths at the attractor locus are given by
%
\begin{equation}\label{eq:backgroundfieldsBH}
 R^2\,F^A= p^A{}' \omega_{\mathbf{S}^2} - 2\text{Re}\, CX^A \omega_{\rm{AdS}_2}\, ,\qquad R^2\, G_A= q_A' \omega_{\mathbf{S}^2} - 2\text{Re}\, C\mathcal{F}_A \omega_{\rm{AdS}_2}\, , 
\end{equation}
%
\end{comment}
% 
Using this, the last term of \eqref{eq:BPSwordlineactionAdS2xS2} can be written as 
\begin{equation}\label{eq:DyonicInteraction}
    \frac{1}{2}\int_\Sigma p^AG_A-q_A F^A = -q_e \int_\gamma \frac{dt}{\rho}-q_m \int_\gamma \cos \theta d\phi \,,
\end{equation}
with 
\begin{equation}\label{eq:chargeDefSugra}
    q_e= \text{Re}\, ( \bar{Z}_{\rm BH} Z_{\rm p} )\, ,\qquad q_m= \text{Im} \,  (\bar{Z}_{\rm BH} Z_{\rm p}) = \frac12 (p^Aq_A'-q_A p^A{}') \,.
\end{equation}
Notice that the parameters $q_e$ and $q_m$ have a clear physical interpretation. The interaction between the BPS black hole and the probe particle is dyonic, such that $q_e$ and $q_m$ then measure, respectively, the effective Coulomb-type interaction (electric-electric and magnetic-magnetic) and the effective mixed interaction (electric-magnetic). For BPS particles, the following identity moreover holds \cite{Castellano:2025yur}
\begin{equation}\label{eq:ChargeMassIdentity}
    m^2 R^2 = |Z_{\rm p} \bar{Z}_{\rm BH}|^2  = q_e^2 + q_m^2 \ge q_e^2  \,,
\end{equation}
which implies that the gravitational attraction that a generic particle feels is stronger than its Coulomb interaction.\footnote{Still, the particles admit trajectories which are supersymmetric, in the sense of $\kappa$-symmetry. See below for details.} Consequently, BPS particles are effectively sub-extremal, in the sense of the (flat-space) Weak Gravity Conjecture \cite{Arkanihamed:2006dz}. Observe that from this perspective, we can already explain why the situation with $\text{Im}(\alpha) = 0$ identified in the previous section is special. In these backgrounds the D0s have $q_m = 0$ and experience a cancellation of forces. The cases with $\text{Re}(\alpha) = 0$ are instead those where $q_e = 0$ and classically there is no Coulomb-type interaction. This latter observation however does not explain the difference between the $q_e = 0$ and the generic subextremal cases, hence suggesting that to shed light on the physics encoded in the non-perturbative corrections to the entropy we need to go beyond the semiclassical properties of the probe branes. In what follows, we will provide additional evidence that these interpretations are the correct ones.

\medskip

We now study the on-shell trajectories of \eqref{eq:BPSwordlineactionAdS2xS2} explicitly. Rewriting the above worldline theory in an equivalent way using an einbein field, and extracting the associated Hamiltonian, we obtain
\begin{equation}\label{eq:worldlineHamiltonian}
  H= \frac12\rho^2 
\left[ p_\rho^2 - \left( p_t + \frac{q_e}{\rho} \right)^2\right] + \frac12p_\theta^2 + \frac12\csc^2\theta\, \left( p_\phi+q_m \cos \theta\right)^2\,,
\end{equation}
where $p_i = \frac{\partial L}{\partial \dot{x}^i}$ are the conjugate momenta. The einbein formulation requires the Hamiltonian constraint $2 H = \tilde{m}^2$ with $\tilde{m} = m R$. The motion of the particles turns out being completely integrable, giving rise to simple trajectories which correspond to circular orbits in $\mathbf{S}^2$---at a fixed polar angle $\theta$---as well as certain (branches of) hyperbolae within AdS$_2$. To show this explicitly, we first note that, since both the electric and magnetic fields are constant and orthogonal to the corresponding 2d surfaces, the physical system inherits the symmetries exhibited by the underlying four-dimensional spacetime \cite{Comtet:1986ki, Dunne:1991cs, Pioline:2005pf}. These are encoded into the (super-)conformal group ${SU(1,1|2)}$, which contains $SU(1,1)$ and $SU(2)$ as bosonic subgroups. The first factor indeed corresponds to the conformal isometries of AdS$_{2}$, generated by $\{K_0, K_{\pm}\}$, whilst the second one captures the rotational symmetry of the sphere, whose generators we denote in the following by $\{J_0, J_{\pm}\}$ (see Appendix \ref{app:detailsTrajectories} for details). One can prove that \eqref{eq:worldlineHamiltonian} may be written as
\begin{equation}\label{eq:algebraicconstraint}
  2H= \delta^{ik} J_i J_k + K_0^2 - \frac12 \left( K_+K_- + K_-K_+ \right) -q_e^2-q_m^2= C_2^{\mathbf{S}^2} + C_2^{\text{AdS}_2} -R^2 |Z_p|^2\,,
\end{equation}
and the Hamiltonian constraint reads
\begin{equation} \label{eq:hamiltonianConst}
  C_2^{\mathbf{S}^2} + C_2^{\text{AdS}_2}=- \Delta^2\, ,
\end{equation}
with $\Delta^2= \tilde{m}^2-q_e^2-q_m^2 \geq 0$. The strategy now is to solve first the sphere dynamics and, subsequently, consider that associated to 2d Anti-de Sitter space, taking into account the mass shell constraint. We observe that the generalized angular momentum $p_\phi = j$ corresponds to just one component (that along the $x^3$-direction) of the three-dimensional vectorial conserved quantity $\boldsymbol{J}$. The latter can be identified with the total angular momentum of the system, i.e., including that associated to the gauge field. If we choose our coordinate system such that the vertical direction is perfectly aligned with $\boldsymbol{J}$, namely we set $J_1=J_2=0$ and $J_3=j$, then the charged particle remains for all times at a certain polar angle $\theta$ 
\begin{equation}
 p_\theta= \pm i \left( \cot \theta\, p_\phi+q_m \csc \theta\right) =0 \iff \cos \theta = -\frac{q_m}{j}\, .
\end{equation}
We can then use the algebraic formulation to solve the charged geodesic equations for the AdS$_2$ factor in an implicit way \cite{Pioline:2005pf}. The constraint \eqref{eq:algebraicconstraint} can be massaged into
\begin{equation}\label{eq:hyperbolaetrho}
  \left( \rho+\frac{q_e}{K_+}\right)^2-\left( t-\frac{K_0}{K_+}\right)^2= \frac{\tilde{m}^2+\ell^2}{K_+^2}\,,
\end{equation}
%
%%%%%%%%%%%%%%%%%%%%%%%%%%%%%%%%%%%%%%%%%%%%%
\begin{figure}[t!]
\centering
\subcaptionbox{ \label{sfig:supextremaltraj1}}{\frame{\includegraphics[width=0.31\linewidth]{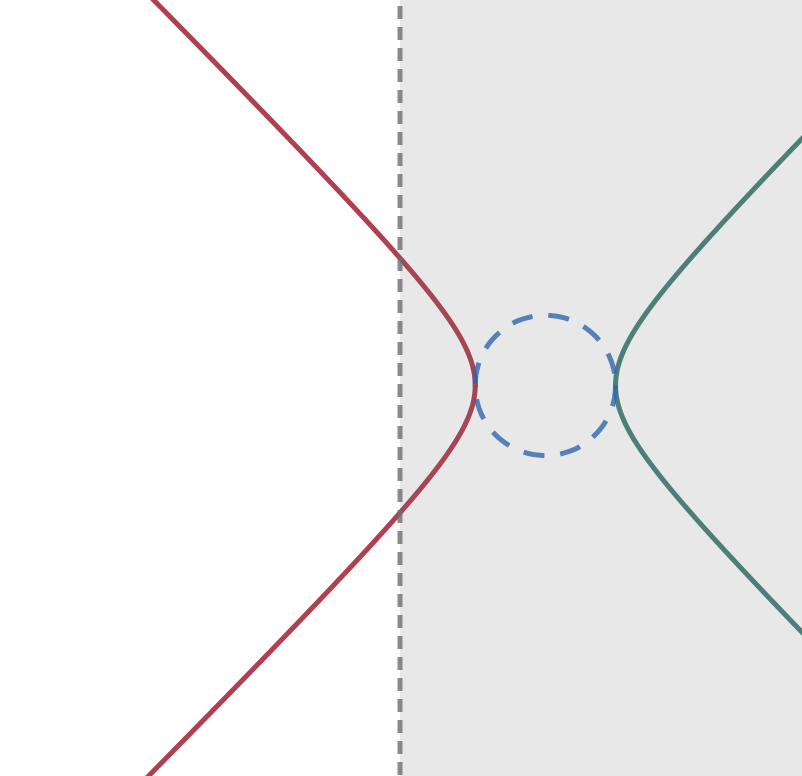}}}
\hspace{0.5cm}
\subcaptionbox{\label{sfig:supextremaltraj2}}{\frame{\includegraphics[width=0.31\linewidth]{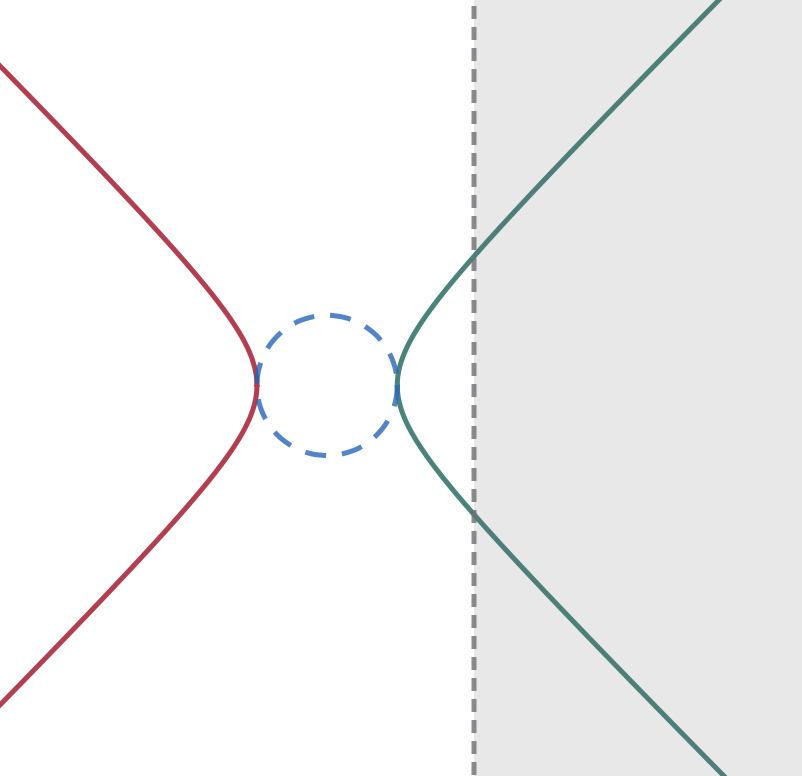}}}
\vspace{0.5cm}

\bigskip

\subcaptionbox{\label{sfig:subextremaltraj}}{\frame{\includegraphics[width=0.31\linewidth]{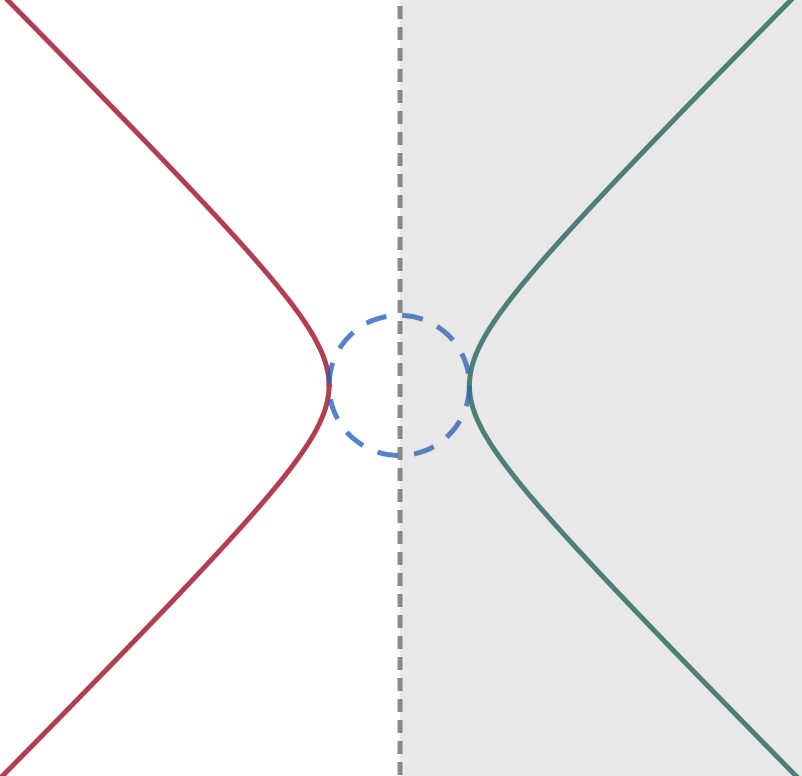}}}
\hspace{0.5cm}
\subcaptionbox{\label{sfig:extremaltraj}}{\frame{\includegraphics[width=0.31\linewidth]{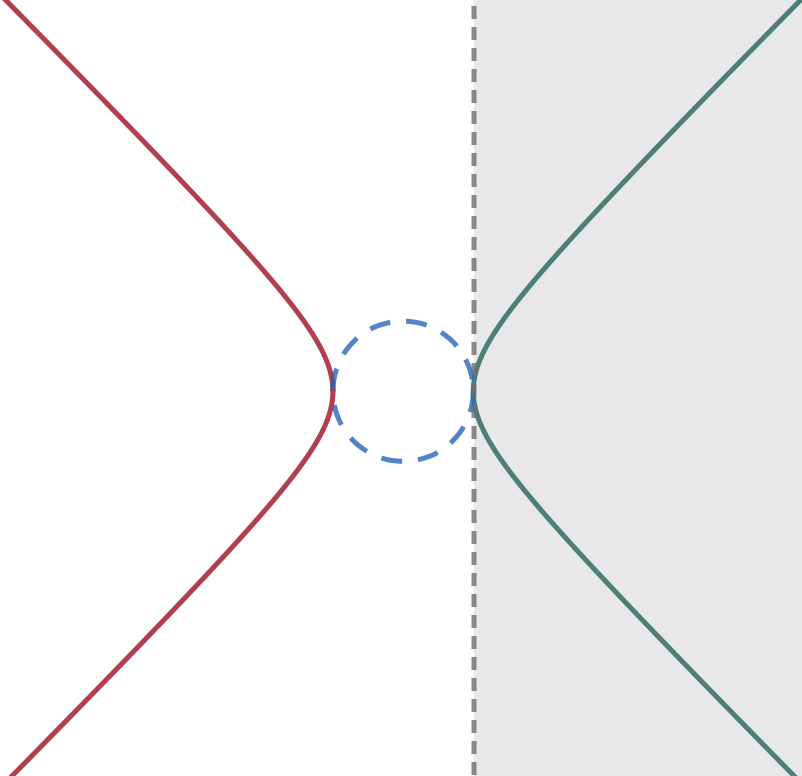}}}
\medskip
\caption{\small Classical paths of charged particles in the Poincaré patch of AdS$_2$ (shaded grey region). The solid green (red) line represents the D0-brane (resp., anti-D0-brane) trajectory parametrized by real proper time, while the dashed blue curves correspond to those in imaginary time. $\textbf{(a)}$ For $\bar{m} < |q_e|$, charged particles with $q_e p_t < 0$ can undergo pair production in the bulk. $\textbf{(b)}$ In contrast, configurations with $q_e p_t > 0$ do not allow for such effects. $\textbf{(c)}$ For $\bar{m} > |q_e|$, the classical trajectories remain confined at a finite radial distance from the boundary, signaling vacuum stability. This corresponds to BPS particles with non-zero magnetic charge, i.e., $q_m \neq 0$. $\textbf{(d)}$ Finally, when $\bar{m} = |q_e|$, the anti-D0-brane trajectory asymptotically approaches the AdS boundary and disappears, while the one of the D0-brane becomes tangent to the latter. This special case corresponds to a BPS particle with $q_m = 0$.}
\label{fig:trajectories}
\end{figure}
%%%%%%%%%%%%%%%%%%%%%%%%%%%%%%%%%%%%%%%%%%%%%%
%
thus leading to hyperbolic trajectories in Poincaré coordinates. The dynamics in the simple case with no orbital angular momentum ($\ell = 0$) is summarized in Figure \ref{fig:trajectories}. The explicit study of geodesics confirms that particles which are subextremal, i.e., satisfy the bound \eqref{eq:ChargeMassIdentity}, are classically confined in the AdS$_2$ throat. Super-extremal particles, namely those having $q_e > \tilde{m}$, can escape. Finally, extremal particles with $\tilde{m}^2 = q_e^2$ admit trajectories that are tangent to the AdS$_2$ boundary, signaling a force cancellation at asymptotic infinity.  

\smallskip

We conclude the analysis of the classical trajectories by commenting on an interesting point. BPS particles which do not saturate the bound \eqref{eq:ChargeMassIdentity} and feel a force, can still produce supersymmetric configurations.
A wrapped D-brane propagating in a \emph{fixed} bosonic background breaks all the spacetime supersymmetries (see \cite{Simon:2011rw} for a comprehensive review). Indeed, a spacetime supersymmetry with Killing spinor $\epsilon$ will always induce a super-translation of the Grassmann-valued superspace coordinates $\delta_\epsilon \Theta = \epsilon $. However, the worldline theory might admit an additional gauge freedom called $\kappa$-symmetry \cite{Becker:1995kb, Bergshoeff:1997kr}, which acts on the fermions as a half-rank projection $\delta_\kappa \Theta = (\mathbf{1} + \Gamma) \kappa$. Here, $\kappa$ is a local Grassmann parameter whereas $\Gamma$ defines a traceless involution that depends on the details of the theory under consideration. Supersymmetry is then restored if the aforementioned global transformation parametrized by $\epsilon$ can be compensated via some $\kappa$-variation. One can verify that BPS particle probes carrying non-vanishing angular momentum ($\ell$) admit supersymmetric worldlines at constant radius in Anti-de Sitter and fixed polar angle on the sphere \cite{Castellano:2025rvn}
\begin{equation}\label{eq:susystationary}
\sinh \chi = \frac{q_e}{|j|}\,,\qquad \cos \theta=-\frac{q_m}{j}\, ,\qquad \frac{d\phi}{d\tau}=\pm1\, ,
\end{equation}
%
%%%%%%%%%%%%%%%%%%%%%%%%%%%
\begin{figure}[t!]
    \centering
\subcaptionbox{\label{sfig:static}}{
		\includegraphics[scale=0.4]{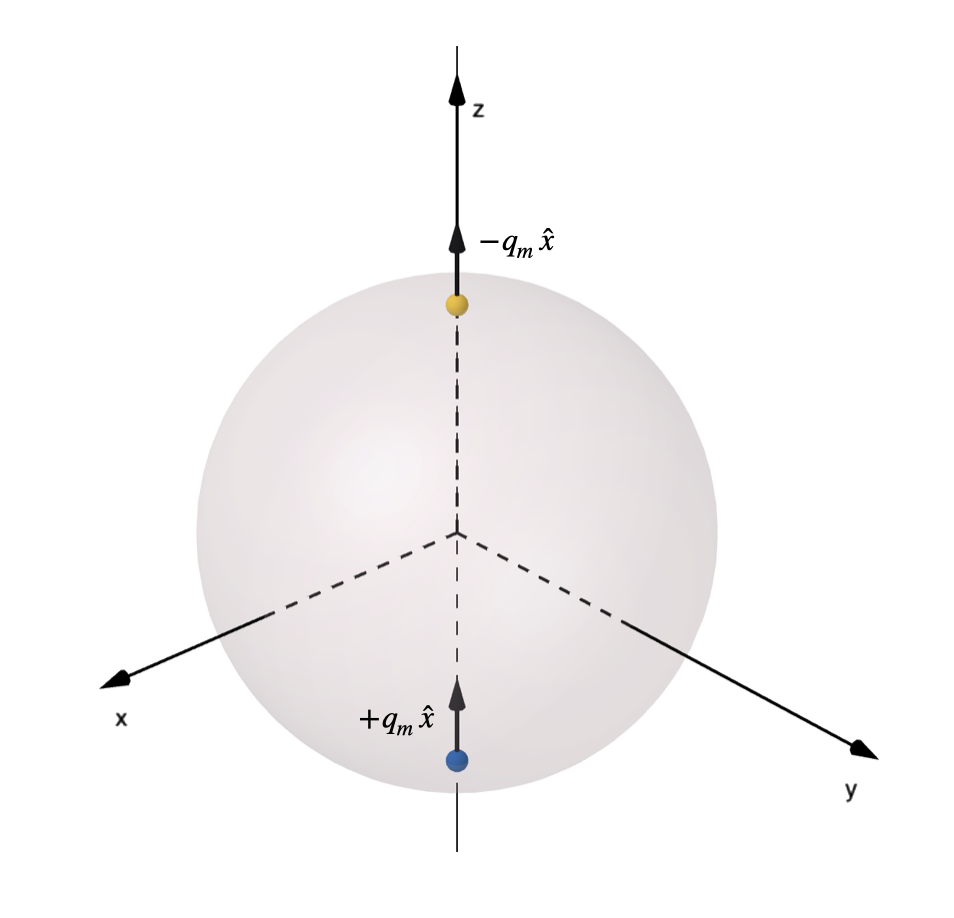}}
        \hspace{0.6cm}
        \subcaptionbox{\label{sfig:stationary}}{\includegraphics[scale=0.35]{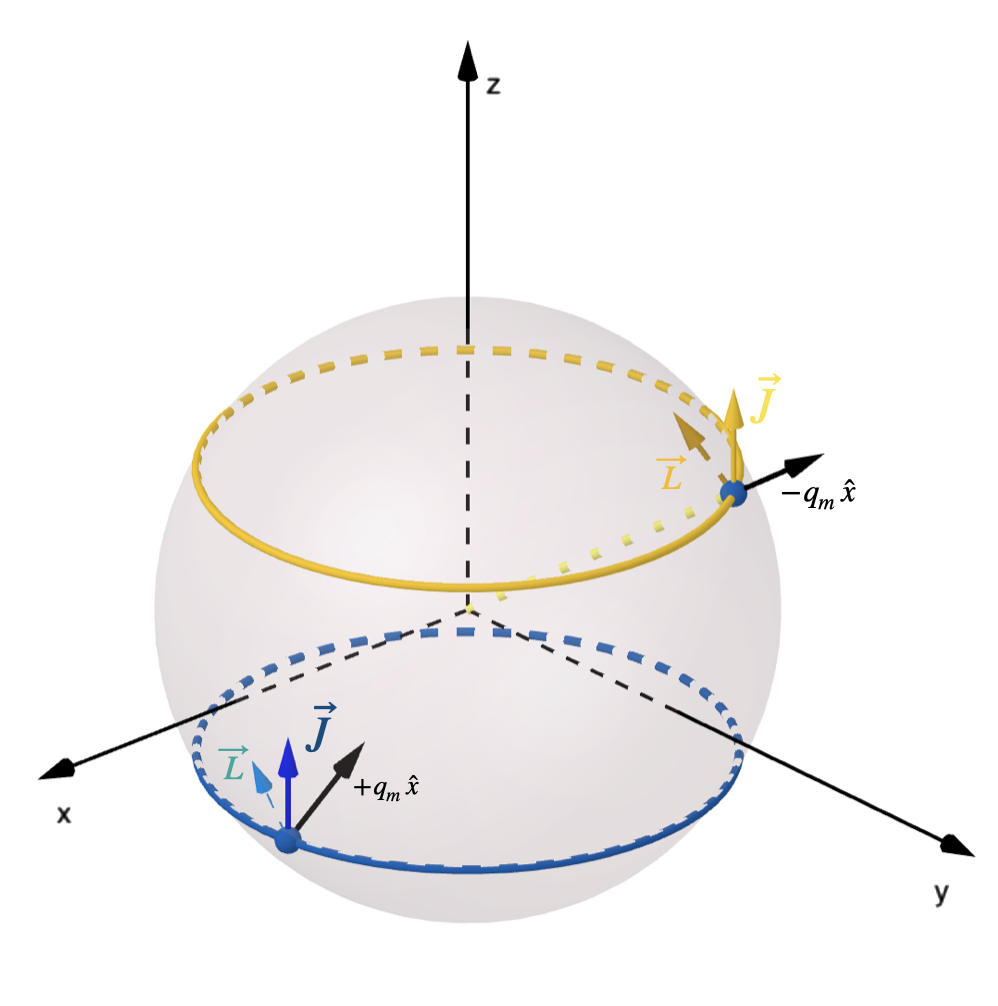}}
		\caption{\small A system comprised by a particle/anti-particle pair can be BPS if the total generalized angular momentum satisfies $|\boldsymbol{J}_{\rm tot}| = |\boldsymbol{J}_1| + |\boldsymbol{J}_2|$. $\textbf{(a)}$ Static configuration with the probes located at antipodal points on $\mathbf{S}^2$. $\textbf{(b)}$ Stationary case with the particles rotating in opposite directions.} 
		\label{fig:multiparticleBPS}
\end{figure}
%%%%%%%%%%%%%%%%%%%%%%%%%%%
where we used global coordinates $ds^{2} = R^2 ( - \cosh^2\chi\, d\tau^{2} + d\chi^2 + d\Omega_{2}^{2})$. In addition, one can show that the direction of the generalized angular momentum vector $\boldsymbol{J}$ specifies the subset of unbroken supercharges, thus allowing for a richer spectrum of multi-particle BPS configurations where all the individual constituents satisfy \eqref{eq:susystationary}, with their corresponding angular momenta being perfectly aligned, see Figure \ref{fig:multiparticleBPS}. This includes, in particular, situations where both particles and anti-particles are present and remain stationary, as opposed to what happens in 4d Minkowski space \cite{Simons:2004nm, Castellano:2025rvn}.

\medskip 

Let us focus now on the purely quantum effects associated with the probe brane. Among them there is Schwinger pair production \cite{Schwinger:1951nm} which may be understood as a quantum tunneling process between classically allowed motions which are nevertheless separated from each other by a finite potential barrier (see e.g., \cite{Brezin:1970xf, Affleck:1981bma, Affleck:1981ag,Coleman:1985rnk,Garriga:1993fh} for an incomplete list of references). The possibility of tunneling is tied to the existence of worldline instanton solutions in the Euclidean theory \cite{Comtet:1986ki}. These give rise, eventually, to an imaginary part in the Wilsonian effective action, which is responsible for the quantum non-persistence (i.e., decay) of the vacuum under consideration. The Euclidean version of the worldline action \eqref{eq:BPSwordlineactionAdS2xS2} restricted to paths which solve the classical equations of motion on the sphere is
\begin{equation}\label{eq:BPSwordlineEuclidean}
 S_{E} = \frac12 \int d\sigma\left[ \left(\frac{1}{h \rho^2}\left(\dot{t}_E^2+\dot{\rho}^2\right)\right)+hm_{\rm eff}^2\right] + q_e\int d\sigma\, \frac{\dot{t}_E}{\rho}\, ,
\end{equation}
where we have neglected the purely magnetic coupling since it does not contribute to the integral, and we moreover defined an effective mass of the form
\begin{equation} \label{eq:effectiveMass}
    m_{\rm eff}=\sqrt{\tilde{m}^2+\ell(\ell+1)+\frac14} \,.
\end{equation}
Here, $\ell(\ell+1)$ originates from the quantization of the sphere spectrum and the additional $\frac14$ factor accounts for the zero point energy of spin-0 particles in AdS$_2$. It is not difficult to verify that the radial equation of the theory \eqref{eq:BPSwordlineEuclidean} takes the form $p_\rho = V(\rho)$  and admits classical solutions connecting the turning points $\rho_{\pm}$ of the potential. For the case where $|q_e| \ge m_{\rm eff}$, the trajectories correspond to circles in the hyperbolic $(t_E,\rho)$-plane \cite{Comtet:1986ki}, as depicted in Figure \ref{fig:trajectories}. The Euclidean action one gets for the corresponding worldline instantons is
\begin{equation}\label{eq:supextremeinstantonaction}
 S_E= 2\pi \left( q_e-\sqrt{q_e^2-\tilde{m}^2-\ell(\ell+1)-\frac14}\right)\, ,
\end{equation}
in agreement with the results of \cite{Pioline:2005pf,Chen:2012zn}. Their precise contribution to the 2d effective action can be obtained upon performing a saddle point approximation of the one-loop effective action
\begin{equation}\label{eq:onelooppathint}
\begin{split}
   S^{\rm 2d}_{\rm eff}[A] & =-\int_0^\infty \frac{dh}{h}\int \mathcal{D} t_E(h)\mathcal{D} \rho(h)\ e^ {-S_{\rm E}[x^ i, A]}\, \\[2mm]
   & \sim\, i\int d^2x\,\sqrt{g}\,\sum_{k=1}^\infty\sum_{\ell=0}^{\ell_{\rm max}} f\left(q_e^2-m_{\rm eff}^2\right)\, e^{-2\pi k \left( q_e-\sqrt{q_e^2-m_{\rm eff}^2}\right)}\,+\, \ldots\, .
\end{split}
\end{equation}
Conversely, when $|q_e| < m_{\rm eff}$ the action is divergent. This reflects the fact that the underlying trajectories become now unbounded, eventually reaching the boundary of $\mathbb{H}^2$. This implies, in turn, the absence of real instanton solutions with finite action in the Euclidean theory, which could be responsible for mediating any type of Schwinger-like vacuum decay. Finally, in the extremal case one similarly obtains a finite answer, namely $S_E=2\pi q_e$, which may be retrieved directly  from eq. \eqref{eq:supextremeinstantonaction} by taking the limit $|q_e| \to m_{\rm eff}$. However, the prefactor accompanying each of those terms---associated with one-loop fluctuations around the saddle point---exhibits a functional dependence on the quantity $q_e^2-m_{\rm eff}^2$ that exactly vanishes for extremal particles \cite{Lin:2024jug}.

\medskip

Let us briefly comment on what we have learned from the semiclassical analysis. First, the classical bound \eqref{eq:ChargeMassIdentity} is not equal to the one relevant for Schwinger pair production. The latter involves the effective mass \eqref{eq:effectiveMass} and the BPS  particles do not satisfy it. Second, we can give a physical interpretation to $\text{Re}(\alpha) = q_e= 0$ black holes as those for which the probe D0-brane cannot be pair produced by the purely magnetic background. Third, one realizes that the contribution of worldline instantons with different values of the angular momentum along the sphere need to be considered (whenever $q_e > m_{\rm eff}$). This reflects the fact that the theory is four-dimensional in origin and we are expanding the contribution of a 4d superextremal particle in spherical harmonics. We conclude by saying that the present semiclassical analysis is thus able to provide quantitative description of some of the features of the black hole entropy. However, it cannot explain by itself their precise form and therefore an exact quantum path integral in AdS$_2\times \mathbf{S}^2$ becomes necessary. This is what we turn to next.

\section{A Path Integral Assessment}\label{s:PathIntegral}

\noindent In chapter \ref{s:4dBHs&NonPerturbativeEffects} a compelling picture emerged. The structure of the corrections to the BPS black hole entropy computed in the fully backreacted theory can be related to properties of probe D-branes in the near-horizon geometry. In this section, we go beyond the semiclassical analysis presented in Section \ref{sec:semiclassicalAnalysis} and determine all relevant quantum effects that massive particles induce to the 4d effective action. More precisely, we compute the exact functional determinants of charged, massive spin-0 and spin-$\frac12$ fields in $\text{AdS}_2\times \mathbf{S}^2$ backgrounds threaded by constant electric and magnetic fields. We then specialize our analysis to supersymmetric settings and we obtain the effective action for a 4d $\mathcal{N}=2$ BPS massive hypermultiplet in $\text{AdS}_2\times \mathbf{S}^2$ attractor geometries. Interestingly, we find that certain combinations of the 1-loop determinants exhibit the same structure of the Gopakumar-Vafa integral \eqref{eq:I(alpha)}, which was the building block of the topological string theory amplitude in flat space.

\subsection{Integrating out charged massive particles in $\text{AdS}_2 \times \mathbf{S}^2$}

\noindent In order to integrate out minimally coupled spin-0 and spin-$\frac12$ particles we employ Schwinger proper-time formalism, which allows us to derive the full \emph{non-perturbative} effective action in the 1-loop and constant background field approximations.

\medskip

For bosons, we consider the 4d Lorentzian quadratic action
\begin{equation}\label{eq:QuadraticScalarAction}
    S[\varphi,\phi] = S[\varphi] + \int d^4x \sqrt{-\det g} \, \phi^\dagger \left[ -\mathcal{D}^2 - m^2 \right] \phi \,,
\end{equation}
where $\phi$ is a complex massive scalar, $\varphi$ is used to collectively denote all the remaining fields, and $\mathcal{D}^2$ is defined in \eqref{eq:opsS2&AdS2}. Using standard methods one can relate the contribution to the exact, Lorentzian 1-loop effective action $\Gamma[\varphi]$ with the trace of the heat kernel operator associated with $\mathcal{D}^2$
\begin{equation}
    \log \mathcal{Z}_\phi := - i \Delta\Gamma[\varphi] =  - \int_{0}^\infty \frac{d\tau}{\tau}  \, e^{-\frac{\epsilon^2}{4\tau}} \, e^{-\tau m^2} \Tr \left[ e^{- \tau  \mathcal{D}^2} \right] \,,
\end{equation}
where $\epsilon > 0$ is a UV regulator. Moreover, noticing that the kinetic operator splits in commuting parts $[\mathcal{D}^2_{{\rm AdS}} ,\mathcal{D}^2_{\mathbf{S}^2}] = 0$, one can reduce the four-dimensional problem into a couple of independent 2d ones, hence decomposing the four-dimensional trace as follows
\begin{equation} 
      \log \mathcal{Z}_\phi = - \int_{0}^\infty \frac{d\tau}{\tau}  \, e^{-\frac{\epsilon^2}{4\tau}} \, e^{-\tau m^2} \mathcal{K}^{(0)}_{\rm AdS}(\tau) \, \mathcal{K}^{(0)}_{\mathbf{S}^2}(\tau)  \,.
\end{equation}
Exploiting the results of Appendix \ref{app:spectrumA2S2} and performing an Hubbard-Stratonovich (HS) transformation \cite{Stratonovich1957OnAM,Hubbard:1959ub} of the AdS$_2$ and $\mathbf{S}^2$ traces, the proper time dependence factorizes in a convenient way and one obtains the integral representation\footnote{Notice the cancellation of the zero point energies of AdS$_2$ and $\mathbf{S}^2$, which only happens when the two radii are equal.} 
\begin{equation}\label{eq:fullintegralspin0}
    \log \mathcal{Z}_\phi = -\int_0^{\infty}  \frac{d\tau }{\tau^2}\, e^{-\frac{\epsilon^2+t^2 +u^2}{4\tau}}\, e^{-\tau \Delta^2} \left[ \frac{V_{\text{AdS}}}{(4\pi R)^2} \int_{\mathbb{R} + i\delta_t} dt\,  W_B(t)  \right] \left[  \int_{\mathbb{R} + i\delta_u} du \, f_B(u) \right]\,,
\end{equation}
where $\Delta^2 = \tilde{m}^2 - e^2 - g^2$, $\delta_t > 0$ and $\delta_u > 0$ are regulators introduced by the HS transform, and 
\begin{equation}
    {f}_B(u) = \frac{d}{du} \left(\frac{e^{i g u}}{\sin\left(\frac{u}{2}\right)}\right) \,, \qquad  {W}_B(t) = \frac{d}{dt}\left( \frac{\cos(et)}{\sinh\left(\frac{t}{2}\right)} \right)\,. 
\end{equation}
The expression \eqref{eq:fullintegralspin0} can be simplified upon performing first the proper-time integral 
\begin{equation}\label{FirstSintegral}
    I_S = \frac{1}{4\pi} \int_0^{\infty} \frac{d\tau }{\tau^2}\, e^{-\frac{\epsilon^2+t^2 +u^2}{4\tau}}\, e^{-\tau \Delta^2} = \frac{\Delta}{\pi} \frac{1}{\sqrt{\epsilon^2 + t^2 + u^2}}\, K_{1}(\Delta \sqrt{\epsilon^2 + t^2 + u^2})\,,
\end{equation}
where $K_1(z)$ is the modified Bessel function of degree 1. Notice that so far the analysis applies to the near-horizon geometry of a generic asymptotically flat, charged, static, and extremal black hole in four dimensions. This last integral thus provides important information about the stability of extremal non-BPS black holes. In particular, $\Delta^2$ plays the same role of the parameter $m_{\rm eff}^2-q_e^2$ introduced in Section \ref{sec:semiclassicalAnalysis}, and it knows about relative strength of the Coulomb-type and gravitational interactions. As long as $\Delta^2 > 0$, $\Delta \in \mathbb{R}$ and hence $K_{1}$ takes real values. When $\Delta^2 < 0$, $\Delta$ becomes purely imaginary and $K_{1}$ develops an imaginary part, signaling that we might have an instability. Still, one needs to verify whether the effective action as a whole develops such an imaginary part. A more systematic analysis of the 1-loop determinant for superextremal particles in AdS$_2 \times \mathbf{S}^2$ is postponed to \cite{WGCandSPP}, where the precise decay condition will be presented.

\smallskip

One can then restrict to 4d $\mathcal{N} = 2$ BPS black holes setting $e = q_e$ and $g = q_m$, with $q_e$ and $q_m$ defined in equation \eqref{eq:chargeDefSugra}. For BPS particles, one has $\Delta = 0$ and we can evaluate exactly $I_S$
\begin{equation} \label{eq:IsPoles}
     I_S = \frac{1}{\pi} \frac{1}{\epsilon^2 + t^2 + u^2} \,.
\end{equation}
The 1-loop determinant \eqref{eq:fullintegralspin0} can be further simplified by noticing that the line integral in $u$ reduces to a contour integral around the simple pole of \eqref{eq:IsPoles} lying in the upper complex plane. Furthermore, thanks to parity considerations in the integral over $t$, one finally obtains the compact formula
\begin{equation} \label{SintegralBoson}
     \log \mathcal{Z}_\phi = -\frac{V_{\text{AdS}}}{2\pi R^2}   \int_{0^+ }^{\infty} \frac{dt}{t} \, W_B(t) f_B(i t) \,.
\end{equation}

\medskip

The computation for fermions follows essentially the same route. We briefly comment on a few technical differences; readers not interested in these details may skip to the next subsection. We start from the Lorentzian action\footnote{\label{fnote:conventionDirac}Our convention is to use the mostly plus signature, with the Dirac matrices satisfying $\{ \gamma^\mu, \gamma^\nu\}=g^{\mu \nu}$, yielding anti-hermitian $\gamma^0$ and hermitian $\gamma^i$. The fermionic kinetic operator reads $\slashed{\nabla} - i \slashed{A} - m $, and $\bar{\Psi} =  i \Psi^\dagger \gamma^0 $.}
\begin{equation}\label{eq:FermionAction}
    S[\varphi,\Psi] = S[\varphi] + \int d^4 x \sqrt{-\det g}\, \bar{\Psi} (i \slashed{D} - m) \Psi  \,,
\end{equation}
where $\Psi$ is a Dirac spinor, $\varphi$ collectively denotes all other dynamical fields in the theory, and $\slashed{D}$ is defined in \eqref{eq:opsS2&AdS2}. Due to the chirality of the kinetic operator, the following identity holds
\begin{equation}
    \det (i \slashed{D} - m)^2  = \det (\slashed{D}^2 + m^2) \,,
\end{equation} 
and we may express the fermionic 1-loop determinant in terms of the heat kernel trace of $\slashed{D}^2$. Again, we can reduce the four-dimensional spectral problem to a couple of two-dimensional ones because  $\slashed{D}$ splits into anti-commuting operators $\left\{\slashed{D}_{\rm AdS} \otimes \sigma^3, \mathds{1} \otimes \slashed{D}_{\mathbf{S}^2}\right\} = 0$, yielding
\begin{equation} 
      \log \mathcal{Z}_\Psi := - i \Delta\Gamma[\varphi] = \frac{1}{2} \int_{0}^\infty \frac{d\tau}{\tau}  \, e^{-\frac{\epsilon^2}{4\tau}} \, e^{-\tau m^2} \mathcal{K}^{(1/2)}_{\rm AdS}(\tau) \,\mathcal{K}^{(1/2)}_{\mathbf{S}^2}(\tau)\,.
\end{equation}
Proceeding as in the bosonic case, we obtain that $\log \mathcal{Z}_\Psi$ is (minus) two times \eqref{SintegralBoson}, with the functions $f_B$ and $W_B$ replaced by 
\begin{equation}
        f_F(u) = \frac{d}{du} \left[  \frac{e^{i g u}}{\tan\left(\frac{u}{2}\right)}\right] \,, \qquad    {W}_F(t) = \frac{d}{dt}\left( \frac{\cos(e t)}{\tanh\left(\frac{t}{2}\right)} \right)\,. 
\end{equation}
The 1-loop determinant for supersymmetric-like particles in BPS $\text{AdS}_2\times \mathbf{S}^2$ backgrounds then reads
\begin{equation}\label{SintegralFermions}
   \log \mathcal{Z}_{\Psi} = \frac{V_{\text{AdS}}}{\pi R^2} \int_{0^+ }^{\infty} \frac{dt}{t} \, W_F(t) f_F(i t) \,,
\end{equation}
where we have set $e = q_e$, $g = q_m$, as well as $\Delta^2 = 0$.

\subsection{Gopakumar--Vafa intregral and non-perturbative corrections}\label{ss:Susic1LoopMinimal}

\noindent Here we want to show how the Gopakumar-Vafa integral \eqref{eq:I(alpha)} naturally emerges when considering supersymmetric 1-loop determinants built with the on-shell field content of a 4d hypermultplet, i.e., two complex scalar fields---forming an $SU(2)_R$ doublet---and one ($R$-singlet) Dirac fermion.
Combining the expressions \eqref{SintegralBoson} and \eqref{SintegralFermions} obtained in the previous section, one finds \cite{Castellano:2026ojb}
\begin{equation}\label{eq:susyeffaction}
    \begin{split}
    \log{\mathcal{Z}_{\rm hm}} & :=  2\log{\mathcal{Z}_\phi} + \log{\mathcal{Z}_{\Psi}} \\[2mm]
    & \hspace{1mm} = -\frac{V_{\text{AdS}}}{\pi R^2}\, \,\text{Re}\left[  q_e q_m \tan^{-1}\left(\frac{q_e}{q_m}\right)  -  \frac{1}{4}\int_{0^+}^{\infty} \frac{dt}{t}  \frac{e^{i(q_e + i |q_m|) t}}{ \sinh^2\left(\frac{t}{2}\right)} \right]\,.
    \end{split}
\end{equation}

Let us comment on various salient features exhibited by the 1-loop determinant thus obtained. Note that the combination $\beta^{-1} = q_e + i q_m$ is nothing but $\bar{Z}_{\rm BH} Z_p$.\footnote{Notice that we changed the definition of $\beta$ with respect to \cite{Castellano:2026ojb}.} This provides a natural generalization of the coupling $\alpha^{-1}$ already encountered in Section \ref{s:4dBHs&NonPerturbativeEffects} when studying the quantum corrections to the black hole entropy induced by certain higher derivative F-terms in the string theory action. Exploiting this analogy further, one can easily understand why $\beta$ is the relevant coupling of the system under consideration without having to rely on the topological string interpretation. Observe that $\beta$ is related to the black hole radius $R$ and the particle mass $m$ via $|\beta|^{-1} = m R $. Therefore, taking $m^{-1}$ to be the Compton wavelength of the particle, it becomes clear that for $|\beta| \ll 1 $, the curvature and graviphoton corrections become suppressed. We should then identify such a regime as the weak coupling limit of the system, and similarly interpret $\beta$ as the (complexified) expansion parameter in the present background. This is further supported by the observation that the presence of poles in the complex Schwinger $t$-plane implies that the 1-loop determinant \eqref{eq:susyeffaction} can be decomposed into a line integral plus an infinite sum over residues, with the latter having a structure that is precisely non-perturbative in $\beta$.\footnote{\label{fnote:nonpertdef}With this we mean that its formal series around $\beta =0$ vanishes term by term.} Consequently, $|\beta| \to \infty$ would correspond to the strong coupling regime, which formally coincides  with the massless (and thus uncharged) limit. Hence, since the latter decouples from the gauge field, one expects  the resulting path integral to be free of non-perturbative ambiguities. What one finds is that indeed the contribution from the poles disappears,\footnote{The reason for this is somewhat technical. One can still pick them up via a suitable complex contour rotation; however, the residues cancel exactly against the contribution from the half-arc around the origin.} providing further evidence that they truly encode non-perturbative information. Notice also that what we reproduce is not directly \eqref{eq:I(alpha)}, but rather its real part, i.e., the correction entering in the free energy $\text{Im}\, F$. More concretely, specifying $q_{e,m}$ for a D0-brane and assuming $0 \le \theta_\alpha \le \pi/2$, one obtains after a few manipulations
\begin{equation}\label{eq:LinkwithGV}
    \text{Re} \,\int_{0^+}^{\infty} \frac{dt}{t}  \frac{e^{i(q_e + i |q_m|) t}}{ \sinh^2\left(\frac{t}{2}\right)} = \text{Re} \,\int_{0^+}^{\infty}\frac{d\tau}{\tau}  \frac{ e^{ - i 4 \pi^2 n^2 \tau} }{ \sinh^2\left(\pi n \alpha \tau \right)} \,.
\end{equation}

Finally, we note that the 1-loop determinant may be written in a more suggestive way as follows (cfr. Appendix \ref{app:spectrumA2S2})
\begin{equation} \label{eq:AdS2xS2effaction}
\begin{aligned}
    \Delta\Gamma_{\rm hm} [A_\mu] =\, &\frac{1}{(4\pi R^2)^2}\int_{\text{AdS}_2\times \mathbf{S}^2} d^4x \sqrt{-\det g}\, \int_{0^+}^{\infty} \frac{dt}{t} e^{- |B|R^2 t} \left[\frac{ \cos(E R^2 t) }{ \sinh^2\left(\frac{t}{2}\right)}\right]\\
    &-\int_{\text{AdS}_2\times \mathbf{S}^2}\frac{\theta_{\rm eff}}{4\pi^2}\, E\, B\, \omega_{\text{AdS}_2} \wedge \omega_{\mathbf{S}^2}\,, 
\end{aligned}
\end{equation}
where we defined $\theta_{\rm eff}:=\tan^{-1}\left(E/B\right)$. The notation has been chosen to reflect the fact that the second term has a form very reminiscent of a topological $\theta$-term \cite{Castellano:2026ojb}.

\subsection{Integrating out supersymmetric particles in AdS$_2 \times \mathbf{S}^2$}\label{ss:SUSYComputation}

\noindent In Section \ref{ss:Susic1LoopMinimal}, we combined the exact functional traces derived for charged, massive scalar and spinor fields in AdS$_2 \times \mathbf{S}^2$ to obtain the 1-loop effective action induced by a BPS hypermultiplet minimally coupled to the near-horizon background of a supersymmetric black hole in 4d $\mathcal{N}= 2$ supergravity. However, the actual two-derivative Lagrangian contains additional interactions in the form of a Pauli-like coupling between the fermionic degrees of freedom and the graviphoton field. In this section will take into account the effect of such terms.

\medskip

We follow the approach of \cite{Keeler:2014bra} and exploit the superconformal symmetries of AdS$_2\times \mathbf{S}^2$ to set up an on-shell computation that manifestly diagonalizes the bulk interactions.\footnote{Another possible strategy is presented in \cite{Castellano:2026ojb} and consists in the explicit diagonalization of the relevant spin-$\frac12$ massive kinetic operator that is similar to the one performed in \cite{Sen:2012kpz} for the massless case.} Particles propagating in such spacetime organize in irreducible representations of $SU(1,1)\times SU(2)$ and are classified by a pair $(h, j)$ of quantum numbers, where $h$ corresponds to the lowest-weight of the $K_0$ generator of $SU(1,1)$ (hence defining a primary state), whereas $j$ labels the appropriate $SU(2)$ representation. From this, one may construct an infinite tower of `descendants' by acting with $K_+$, thereby raising the $h$ number by one unit each time. Similarly, the allowed values of $j$ depend on the magnetic field $B=q_m/R^2$ threading the 2-sphere, and correlate with the spin $s$ of the field as $j = n + |q_m| - s$, with $n=0,1\ldots, \infty$. The on-shell condition then amounts to a certain constraint satisfied by the quadratic Casimir operators of $SU(1,1)$ and $SU(2)$ \cite{Castellano:2025rvn}, which relates, in turn, $h$ and $j$. When embedded into $\mathcal{N}=2$ supergravity, the background solution preserves 8 supercharges, and the symmetry group enhances to $SU(1,1|2)$. Therefore, the fields defined therein must furnish themselves representations of the superconformal algebra. When the particles are in addition BPS, the representations become `short', since only half of the available supercharges act non-trivially on those. This means that the possible set of $(h, j)$ must organize into different combinations of chiral multiplets, which for us will take the form \cite{Keeler:2014bra}
\begin{equation}\label{eq:chiralmultiplet}
    (j,j) \oplus 2 \times \left( j+\frac12,j-\frac12\right) \oplus (j+1, j-1)\,.
\end{equation}
Since a 4d hypermultiplet contains four bosonic as well as fermionic degrees of freedom, one quickly realizes that there is a unique way to arrange those in terms of (towers of) chiral multiplets. Namely, one finds two copies of the set
\begin{equation}\label{eq:chiralprimariesAdS2xS2}
    \left(k+|q_m|+\frac12, k+|q_m|+\frac12\right) \oplus 2\times \left(k+|q_m|+1, k+|q_m|\right) \oplus \left(k+|q_m|+\frac32, k+|q_m|-\frac12\right)\,,
\end{equation}
with $k\in \mathbb{Z}_{\geq 0}$. Crucially, the reorganization of the different modes within the hypermultiplet in terms of chiral states already diagonalizes all interactions required by $\mathcal{N}=2$ supergravity. Hence, we can read directly from \eqref{eq:chiralprimariesAdS2xS2} the contribution of each of these pieces to the 1-loop path integral. The fermionic piece arises from the first and third terms, and one obtains\footnote{The  contributions due to on-shell states of the form $(h,j)$ and $(h=1, j=0)$ are related by 
\begin{equation}\label{eq:Spin12TraceContributions(h,j)}
    \mathcal{K}^{(1/2)}_{\text{AdS}_2\times \mathbf{S}^2}(h, j;\tau) =  \mathcal{K}^{(1/2)}_{\text{AdS}_2\times \mathbf{S}^2}(h=1, j=0;\tau) e^{-\frac{\tau}{R^2} \left(h(h-1)-q_m^2+\frac14\right)} (2j+1)\,. \notag
\end{equation}
} 
\begin{equation}\label{eq:HeatKernelTraceHyperFermions}
   \mathcal{K}^{(1/2)}_{\text{AdS}_2\times \mathbf{S}^2}(\tau) = \mathcal{K}^{(1/2)}_{\rm AdS}(\tau)\,  2\, e^{ \frac{\tau}{R^2} q_m^2} \left(2\sum_{n \geq 1} \left( n + |q_m|\right) e^{ -\frac{\tau}{R^2} \left( n + |q_m| \right)^2} + (|q_m|+1)\,e^{ -\frac{\tau}{R^2} q_m^2}\right)\,.
\end{equation}
The Dirac fermion does not reproduce the 1-loop determinant derived for minimally coupled spin-$\frac12$ fields in \eqref{SintegralFermions}. However, it almost does so, with the only difference being captured by the additional `$+1$' in the second term of eq.~\eqref{eq:HeatKernelTraceHyperFermions} above. The latter gives a contribution that is formally equivalent to the presence of two additional zero modes of the Dirac operator on the sphere (cfr. eq.~\eqref{summary:fermionS2}). Repeating the same steps for the scalar sector (middle term of \eqref{eq:chiralprimariesAdS2xS2}) we match precisely twice the result obtained for a charged scalar (eq.~\eqref{eq:fullintegralspin0}). We can then write the complete functional trace associated with a massive BPS hypermultiplet as
\begin{equation}\label{eq:finalhyperformula}
    \log{\mathcal{Z}_{\rm hm}} =\, -\frac{V_{\text{AdS}}}{\pi R^2} \text{Re}\left[ q_e q_m \tan^{-1}\left(\frac{q_e}{q_m}\right)  +  \frac14\int_{0^+}^{\infty} \frac{dt}{t}  \frac{e^{i(q_e + i|q_m|) t} }{ \sinh^2\left(\frac{t}{2}\right)} - i q_e\int_{0^+}^{\infty} \frac{dt}{t} \frac{ e^{i(q_e + i|q_m|) t}  }{ \tanh\left(\frac{t}{2}\right)}\right]\,.
\end{equation}
Note that the $\operatorname{csch}^2(t/2)$ has the structure of the Gopakumar-Vafa integral, but the sign has been flipped with respect to that appearing in \eqref{eq:AdS2xS2effaction}. The third term, on the other hand, is completely new. It can be seen to cancel identically for purely magnetic backgrounds and in principle it modifies both the perturbative and non-perturbative structure of the 1-loop effective action.

\medskip

Let us conclude by commenting briefly on the relevance of this computation. In the setup originally considered by Gopakumar--Vafa, namely Minkowski space with an anti-self dual graviphoton background, the hypermultiplet not only furnishes one of the basic matter contents of the 4d $\mathcal{N}=2$ $\mathbb{R}^{1,1}$ theory but also provides (twice) the minimal representation of the BPS superalgebra \cite{Weinberg:2000cr}. Consequently, any irreducible representation of the supersymmetry and SU(2) massive little groups can be obtained by tensoring the latter with the spin $j$ of the Clifford vacuum, which allow us to reduce its 1-loop contribution to a graded sum over that of the basic building block \cite{Dedushenko:2014nya}. Given this, one might then wonder whether we should expect the same phenomenon to occur in $\text{AdS}_2 \times \mathbf{S}^2$. Despite not having a definite answer, we believe that this might be in fact the case. However, since, as explained in this section, BPS states decompose now into chiral multiplet irreps \eqref{eq:chiralmultiplet}, it is natural to expect these to play a role analogous to that of the hypermultiplet representation in flat space. We plan to explore this point in future work. 

\section{Discussion and Outlook}\label{sec:Outlook}

\noindent In the works reviewed herein \cite{Castellano:2025ljk, Castellano:2025yur, Castellano:2025rvn, Castellano:2026ojb}, we have initiated a more systematic study of the non-perturbative entropy of four-dimensional supersymmetric (single-centered) black holes and its behavior along moduli space trajectories. We studied solutions which can be obtained by wrapping D$2p$-branes, with $p=0,1,2,3$, in Type IIA string theory compactified on Calabi-Yau threefolds. Our starting point was 4d $\mathcal{N}=2$ supergravity effective theory, supplemented with an infinite set of higher derivative F-term-like operators, as reviewed in Section \ref{ss:HigherDerivativeOps}. Crucially, their inclusion is encapsulated by a single holomorphic and homogeneous `prepotential' function $F(X^A, W^2)$ of the vector multiplet moduli and the graviphoton, from which all the relevant black hole observables may be determined. In particular, its imaginary part, when evaluated in the attractor geometry (cfr. eq. \eqref{eq:attractoreqs}), plays the role of the black hole free energy. We focused on class of solutions within the large volume regime (see footnote \ref{fnote:largevol}), where the prepotential can be approximated by tree-level and genus-1 contributions, plus an infinite (asymptotic) series in some perturbative complex expansion parameter $\alpha$ fixed by the attractor mechanism (see eqs. \eqref{eq:holomorphicprepotential@largevol}-\eqref{eq:Gfactors}).

\smallskip

To go beyond the 4d local and perturbative approximation and explore the strong coupling regime in $\alpha$, a non-perturbative definition of the free energy is needed. Interestingly, this lies within our reach thanks to the seminal work of Gopakumar and Vafa \cite{Gopakumar:1998ii, Gopakumar:1998jq}, who showed that the prepotential may equivalently be obtained from the 1-loop determinant of supersymmetric wrapped D-particles in flat space with constant anti-self dual graviphoton background. At large volume, the leading contribution arises from the tower of D0-branes, giving rise to a Schwinger integral of the form shown in eqs. \eqref{eq:G&I(alpha)} and \eqref{eq:I(alpha)}. When expanded around the weak coupling point, $\alpha=0$, this reproduces the asymptotic series, while also extending analytically the result toward larger couplings. This involves a careful evaluation of the Schwinger integral, revealing the presence of infinitely many poles in the complex proper-time plane. The contribution from the series of residues is displayed in \eqref{eq:Inonpertalpha1stmethod}, and it generically leads to a non-perturbative piece in the black hole entropy, in the sense of footnote \ref{fnote:nonpertdef}. The most notable exceptions occur when the expansion parameter $\alpha$ becomes either purely real or purely imaginary, where the absence of such effects may be even crucial to match with microscopic counting results.

\medskip

In order to gain a better understanding of the physics behind all these observations and extract a guiding principle for future generalizations, we studied the dynamics associated with probe D-branes---the same ones giving rise to the prepotential corrections---directly in the near-horizon geometry (cfr. Section \ref{sec:semiclassicalAnalysis}). The latter exhibits some uniform graviphoton background, which is perceived by charged particles through certain effective electric ($q_e$) and magnetic ($q_m$) couplings, defined in \eqref{eq:chargeDefSugra}. Importantly, for BPS states these may be combined into some complex charge $\beta^{-1}= q_e+i q_m$, which is the natural generalization of the coupling constant $\alpha$ controlling the series expansion in the prepotential, and whose modulus is related to the mass of the multiplet in units of the AdS$_2$ radius, namely $|\beta|=(mR)^{-1}$. This has several interesting consequences. First, it endows the coupling parameter of the system with a simple physical meaning, namely as the ratio of the Compton wavelength of the particle to the black hole radius. Thus, for weak curvatures ($|\beta|\ll1$), any correction coming from integrating out the massive D-branes should be suppressed, whereas in the strongly coupled regime ($|\beta| \gg 1$) quantum effects can become important. Secondly, one readily realizes that since $q_e$, which measures Coulomb-type interactions between black hole and probe particle, is at most equal to the D-brane mass, Schwinger pair production can never occur (see discussion around \eqref{eq:BPSwordlineEuclidean}). This renders the AdS$_2 \times \mathbf{S}^2$ geometry---and thus the black hole---non-perturbatively stable, at least against this type of decay channel, making manifest that the residue structure of the Schwinger integral truly accounts for some kind of non-perturbative contributions to the black hole entropy.

\medskip

\begin{figure}[t]
        \centering
        \includegraphics[scale=0.30]{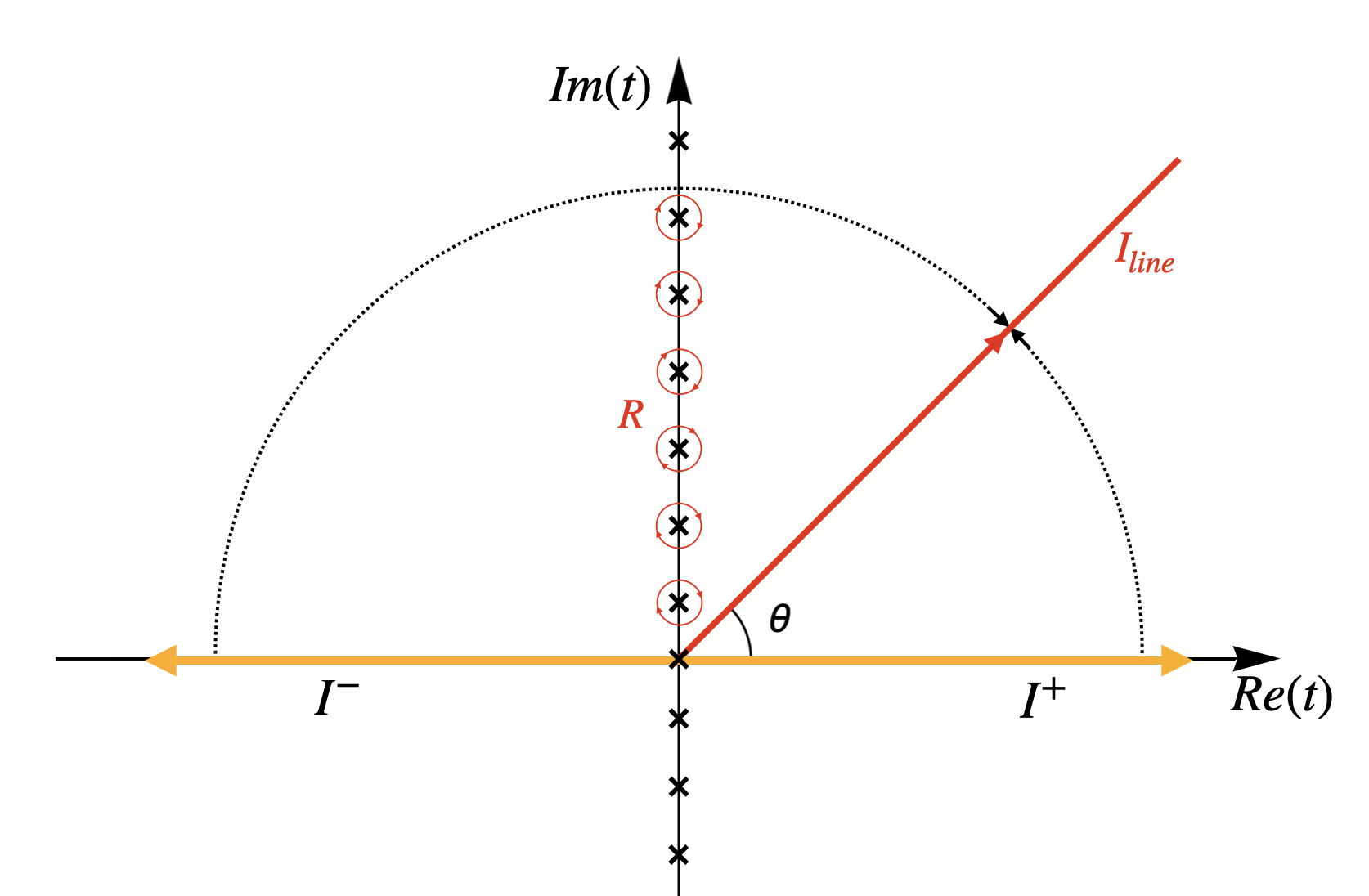}
        \caption{\small To connect \eqref{eq:susyeffaction2} with the Gopakumar-Vafa integral and make manifest the perturbative and non-perturbative structure of the black hole free energy one separates the latter into two and deforms the contours toward the preferred ray at $\theta=\tan^{-1} (|q_m|/|q_e|)$. As a result, one gets a line integral $\mathcal{I}_{\rm line}$ plus some residues.}
        \label{fig:NonPertContour}
\end{figure}

Additional evidence in favour of this picture may be gathered by performing the exact 1-loop path integral computation in the attractor background. Notice that this is somewhat similar to the original Gopakumar-Vafa analysis, albeit in the real geometry that is most relevant for determining the black hole partition function itself. This is the content of Section \ref{s:PathIntegral}, and the result of considering the contribution due to a pair of complex scalar fields and one Dirac fermion---assuming minimal coupling---is shown in \eqref{eq:susyeffaction}, which we repeat here for the comfort of the reader
\begin{equation}\label{eq:susyeffaction2}
    \begin{split}
    \log{\mathcal{Z}_{\rm hm}} = \frac{V_{\text{AdS}}}{4\pi R^2}\, \,\text{Re}\left[ -4q_e q_m \tan^{-1}\left(\frac{q_e}{q_m}\right)  +\int_{0^+}^{\infty} \frac{dt}{t}  \frac{e^{i(q_e + i |q_m|) t}}{ \sinh^2\left(\frac{t}{2}\right)} \right]\,.
    \end{split}
\end{equation}
Let us comment briefly on some features of \eqref{eq:susyeffaction2} which we did not include in Section \ref{s:PathIntegral} given that originally they were not discussed in \cite{Castellano:2026ojb}. The first term has a behaviour very reminiscent of a topological $\theta$-term, with a field-dependent angle given by $\theta_{\rm eff}=\text{arg} (\beta^{-1})-\pi/2$ (see discussion around \eqref{eq:AdS2xS2effaction}). Indeed, it exhibits the correct normalization and is moreover weighted by the number of zero modes of the Dirac operator in AdS$_2\times \mathbf{S}^2$, in accordance with the Atiyah-Singer index theorem \cite{Castellano:2026ojb}.\footnote{Strictly speaking, however, since the background is kept fixed and has constant gravitational and gauge curvatures, it is not obvious a priori whether the resulting term is truly topological.} The second term, on the other hand, can be related with the Gopakumar-Vafa integral, as it was explained for the D0-branes of interest around equation \eqref{eq:LinkwithGV}. To make this connection even more transparent, we repeat here the argument in section 3.3.2 of \cite{Castellano:2026ojb}. Denoting by $\mathcal{I}$ the real part of the integral in \eqref{eq:susyeffaction2}, one can decompose it into a line integral plus a tower of residues according to $  \mathcal{I} =  \mathcal{I}_{\text{line}} +  \mathcal{R} $. To achieve this, we split $  \mathcal{I}$ in half and then change coordinates $t\rightarrow -t$ in one of the contributions, expressing it as an integral over the negative real axis. Next, we align the two contours by performing rotations within the complex $t$-plane. In particular, this can be done ensuring that the arc at infinity does not contribute (cfr. Figure \ref{fig:NonPertContour}), and it also allows us to pick up the residues associated with the integrand at $t= 2\pi i k$ for $k\in \mathbb{N}$, yielding the decomposition
\begin{subequations}\label{eq:deformedIntegral}
\begin{align} 
    \mathcal{I}_{\text{line}} & = \text{Re} \left[\int_{0^+}^\infty \frac{dt}{t}  \frac{ \cos(t) }{ \sinh^2\left(\frac{ \tilde{\beta}\,t}{2}\right)} \right] \,,\label{eq:rotatedLine} \\[2mm]
 \mathcal{R} & = \text{Re} \left[  \, - 2 \pi i\, \sum_{k \ge 1} \text{Res} \, \left(  \frac{ e^{i t/\tilde{\beta}} }{ 2t \sinh^2\left(\frac{t}{2}\right)},\,  2 \pi i k\right) \, \right] = \frac{1}{\pi} \, \text{Im} \left[ \mathrm{Li}_2(e^{-2 \pi/ \tilde{\beta}}) + \frac{2 \pi}{\tilde{\beta}} \mathrm{Li}_1(e^{-2 \pi/\tilde{\beta}})\right]\,, \label{eq:residuesR}
\end{align}
\end{subequations}
where we have defined $\tilde{\beta}^{-1}=|q_e|-i |q_m|$. Summing over an infinite Kaluza-Klein tower (such as the one furnished by the D0-branes) one retrieves exactly \eqref{eq:I(alpha)pert} and \eqref{eq:Inonpertalpha1stmethod}, with $\alpha$ replaced by $\tilde{\beta}$.\footnote{Notice that the cosine in the line integral \eqref{eq:rotatedLine} is crucial for the Poisson resummation of the series. This is the reason why we need to split the initial line integral into half.} Naively then, it would seem that the first term of \eqref{eq:susyeffaction2} gives an extra contribution with respect to the result obtained in Section \ref{sec:D0branesEffects}. However, in the process of deformation, one must recall that we are imposing a UV cutoff $\epsilon \to 0$ in the $t$-integral, thereby requiring special care that has so far been neglected. This thus prompt us to consistently evaluate the contribution due to the small arcs around $t=0$ when following the steps that took us to \eqref{eq:deformedIntegral}, from which we obtain an additional piece that reads
\begin{equation} 
    \mathcal{I}_\epsilon =  - (\pi-2\theta)\, \text{Im}\, \tilde{\beta}^{-2}= 4 q_e q_m \tan^{-1} (q_e/q_m)\,,
\end{equation}
with $\theta=\tan^{-1} (|q_m|/|q_e|)$. Notice that this exactly cancels the $\theta$-term contribution when inserted back into \eqref{eq:susyeffaction2}, hence matching the result of the flat space computation in Section \ref{sec:D0branesEffects}.\footnote{In reality, a similar index-like term also arises in flat space when carefully accounting for the fermion zero modes  \cite{Dedushenko:2014nya}. The latter precisely cancels the semi-arc around the origin that would otherwise contribute to \eqref{contourIntgral}, see Figure \ref{fig:contourIntegralComplex}.}

%%%%%%%%%%%%%%%%%%%%%%%%%%%%%%%%%%%%%
%%%%%%%%%%%%%%%%%%%%%%%%%%%%%%%%%%%%%

\medskip

Despite our findings being very encouraging, we also pointed out various interesting puzzles. Most notably, the analysis of Section \ref{ss:SUSYComputation} shows how the actual supersymmetric computation for the 4d hypermultiplet differs from the minimally coupled result. In particular, integrating out hypermultiplets directly in the AdS$_2 \times \mathbf{S}^2$ geometry is not equivalent to doing so in flat space and, subsequently, evaluate the corrections at the attractor point. More importantly, we observed that the hypermultiplet no longer furnishes the minimal BPS representation of the supersymmetry algebra, unlike what happened in 4d Minkowski. %This apparent mismatch might be related to the fact that in the flat case all supermultiplets produced correctionos proportional to the one of the hypermultiplets. 
Instead, as explained in Section \ref{s:PathIntegral}, these role is now played by the chiral multiplet displayed in \eqref{eq:chiralmultiplet}, in terms of which one may conveniently decompose all BPS representations. We will investigate this further in future work.

\medskip

On another note, our results open several promising directions for future work. A primary motivation for this analysis is the broader goal of understanding the non-perturbative structure of black holes in string theory and quantum gravity. In this context, BPS solutions with AdS$_2 \times \mathbf{S}^2$ near-horizon geometries provide a natural setting for explicit computations. The techniques developed here aim to systematically characterize the effective action induced by massive fields in such backgrounds, with direct implications for black hole physics and quantum entropy functions \cite{Sen:2008yk,Sen:2008vm}, and offer a framework for comparison with existing string theory results \cite{LopesCardoso:1998tkj,LopesCardoso:1999cv,LopesCardoso:1999fsj}. Given the similarity with the original Gopakumar–Vafa construction in flat space, it would be interesting to establish a precise relation between the 1-loop partition function derived here and that framework, clarifying its implications for $\mathcal{N}=2$ single-centered black hole partition functions \cite{Ooguri:2004zv}. Further directions include extending the analysis to other supermultiplets in 4d $\mathcal{N}=2$ theories, including higher-spin states, as well as to other near-horizon geometries in higher dimensions, e.g., AdS$_3 \times \mathbf{S}^2$ or AdS$_2 \times \mathbf{S}^3/\mathbb{Z}_k$. It would also be worthwhile to connect these results with complementary approaches, including microscopic counting and supersymmetric localization techniques \cite{Sen:2007qy,Cassani:2025sim}.

Another promising direction involves the study of family of black holes which allow us to explore different corners of moduli space, beyond the large volume point of Calabi--Yau compactifications. Among these, particularly interesting are the so-called emergent string and F-theory limits \cite{Lee:2019wij}. There, the dominant tower of light states consists not of D0-branes, but rather of a weakly coupled fundamental string or infinitely many D0-D2 bound states, respectively. Such limits admit a perturbative description in terms of a dual heterotic string theory on K3$\times \mathbf{T}^2$ or 6d F-theory compactified on an elliptically fibered Calabi--Yau threefold. Investigating the behavior of non-perturbative corrections in these settings, as well as studying BPS black hole transitions near the aforementioned limits would be of significant interest. These questions are also closely tied to the UV-IR connection between black holes and infinite towers of (light) states, and to the physics of small black hole solutions, where classical supergravity ceases to be reliable and genuine quantum gravity effects may even become dominant. We aim to return to these questions in future work.

\smallskip

We hope that these results provide useful insights into the quantum structure of extremal black holes and serve as a basis for further investigation.

\paragraph{Acknowledgments.} We are grateful to Dieter L\"ust and Carmine Montella for collaboration on some of the results reviewed in this work.~M.Z. acknowledges the hospitality of the Corfu Summer Institute 2025, where part of the research presented here was completed. The work of A.C. is supported by a Kadanoff and an Associate KICP fellowships, as well as through the NSF grants PHY-2014195 and PHY-2412985. The work of M.Z. is supported by the Alexander von Humboldt Foundation.~A.C. and M.Z. are grateful to Teresa Lobo and Miriam Gori for continuous encouragement and support.

\appendix

\section{Two-derivative 4d $\mathcal{N} = 2$ Supergravity} \label{app:4dN2sugra}
 
\noindent We briefly review our conventions, as well as the main features of four-dimensional $\mathcal{N} = 2$ supergravity arising from Type IIA string theory compactified on a Calabi–Yau threefold $X_3$, truncated at the two-derivative level. Our treatment largely follows \cite{Castellano:2025yur}.

The low-energy theory comprises one gravity multiplet, $n_V = h^{1,1}(X_3)$ Abelian vector multiplets, and $n_H = h^{2,1}(X_3) + 1$ massless, neutral hypermultiplets. In what follows, we will concentrate just on the first two, since the black hole solutions we care about only depend on those. Restricting ourselves to the bosonic sector, the latter includes the standard metric, as well as $n_V$ complex scalar fields and $n_V + 1$ gauge bosons. The two derivative action can be written as follows
\begin{equation}\label{eq:IIAaction4d}
		\ S\, =\,  \frac{1}{2\kappa^2_4} \int \mathcal{R} \star 1 + \frac{1}{2}\, \text{Re}\, \mathcal{N}_{AB} F^A \wedge F^B + \frac{1}{2}\, \text{Im}\, \mathcal{N}_{AB} F^A \wedge \star F^B  - 2 G_{a\bar b}\, d z^a\wedge \star d\bar z^b \,.
\end{equation}
with $A = 0, \ldots, n_V$ and $a = 1,  \ldots, n_V$. The kinetic functions associated with the scalar and gauge fields may appear rather complicated; however, the vector moduli space is mathematically described as a projective special K\"ahler manifold \cite{deWit:1980lyi,deWit:1984wbb,deWit:1984rvr,Cremmer:1984hj} which is fully specified by a holomorphic function referred to as the prepotential $\mathcal{F}(X^A)$. The $X^A$ are homogeneous (projective) coordinates on a special K\"ahler manifold and they are related to the lowest spin components $z^i$ of the vector multiplets via $z^a = X^a/X^0$.
The prepotential is moreover homogeneous of degree two, meaning that it satisfies $\mathcal{F}= \frac{1}{2} X^A \mathcal{F}_A$, where $\mathcal{F}_A = \partial_{X^A} \mathcal{F}$. The metric tensor $G_{a\bar b}=\partial_a \partial_{\bar{b}} K$ is related to the Kähler structure moduli via the K\"ahler potential
\begin{equation} \label{eq:kahlerpotential}
K = - \log i\left( \bar{X}^A \mathcal{F}_{A}- X^A \bar{\mathcal{F}}_{A}  \right)\, .
\end{equation}
Similarly, the complexified gauge kinetic function $\mathcal{N}_{AB}$ appearing in \eqref{eq:IIAaction4d} is the expression \cite{Ceresole:1995ca}
\begin{equation}\label{eq:Nab}
    \mathcal{N}_{AB} = \overline{\mathcal{F}}_{AB} + 2i \frac{(\text{Im}\, \mathcal{F})_{AC} X^C (\text{Im}\, \mathcal{F})_{BD} X^D}{X^C (\text{Im}\, \mathcal{F})_{CD} X^D}\, ,
\end{equation}
where \(\mathcal{F}_{AB} = \partial_{X^A} \partial_{X^B} \mathcal{F}\).

\section{Some Details on the Worldline Hamiltonian and Noether Charges} \label{app:detailsTrajectories}

\noindent Let us consider the Hamiltonian \eqref{eq:worldlineHamiltonian}. In terms of the set of canonical variables $\{q^i\}= \{ t, \rho,\theta, \phi\}$ together with their conjugate momenta 
\begin{equation}\label{eq:conjugatemomenta}
p_\rho = \frac{\dot{\rho}}{\rho^2}\, ,\quad
p_t = - \frac{\dot{t}}{\rho^2} - \frac{q_e}{\rho}\, ,\quad
p_\theta =  \dot{\theta}\, ,\quad
p_\phi = \sin^2\theta\, \dot{\phi}-q_m \cos \theta\, .
\end{equation}
which satisfy the usual Heisenberg algebra\footnote{\label{fnote:poissonbracket}The Poisson bracket is defined as
\begin{equation}\label{eq:Poissonbracketdef}
\begin{aligned}
    \big \{ A(q,p), B(q,p) \big\}_{\rm PB}= \frac{\partial A}{\partial q^i}\frac{\partial B}{\partial p_i}-\frac{\partial B}{\partial q^i}\frac{\partial A}{\partial p_i}\, .
    \end{aligned}
\end{equation}
Recall that from the Hamilton equations of motion, $\dot{q}^i= \frac{\partial H}{\partial p_i}$, $\dot{p}_i= -\frac{\partial H}{\partial q^i}$, it follows that the (proper) time evolution of any function $A=A(q^i, p_j)$ is determined by its Poisson bracket with the Hamiltonian, namely
\begin{equation}
\begin{aligned}
    \frac{dA(q,p)}{d\sigma} =\big \{ A(q,p), H(q,p) \big\}_{\rm PB}\, .\notag
    \end{aligned}
\end{equation}}
\begin{equation}\label{eq:Heisenbergalgebra}
\begin{aligned}
    \big \{ q^j,p_k \big\}_{\rm PB}= \delta^j_k\, ,
    \end{aligned}
\end{equation}
the generators of the corresponding symmetry groups are given (in Chevalley form) by
\begin{equation}\label{eq:SL2generators}
\begin{aligned}
    &K_+ = p_t\, ,\\
    &K_- = (t^2 + \rho^2) p_t + 2 t\rho\, p_\rho +2 q_e\rho\, ,\\
    &K_0 = t\, p_t + \rho\,  p_\rho\, ,
    \end{aligned}
\end{equation}
for the $SL(2,\mathbb{R})$ conformal group of AdS$_2$, and similarly
\begin{equation}\label{eq:SU2generators}
\begin{aligned}
    &J_1 = -\sin \phi\, p_\theta-\cot \theta \cos\phi\, p_\phi - q_m \csc \theta \cos \phi\, ,\\
    &J_2 = \cos \phi\, p_\theta-\cot \theta \sin\phi\, p_\phi - q_m \csc \theta \sin \phi\, ,\\
    &J_3 = p_\phi\, ,
    \end{aligned}
\end{equation}
for the rotational $SU(2)$ group associated to the 2-sphere. Note that $K_+$ and $K_0$ have a simple interpretation as (Poincaré) time translation and dilatation rescaling operators, respectively, whereas $K_-$ generates certain non-linear special conformal transformations. With this at hand, one may readily check that these functions satisfy the algebra
\begin{equation}\label{eq:SU2xSL2algebra}
\begin{aligned}
    &\big \{ J_i, J_j \big\}_{\rm PB} = \epsilon_{ijk}J_k\, ,\\
    &\big \{ K_+, K_- \big\}_{\rm PB} = -2K_0\, , \quad \big \{ K_0, K_\pm \big\}_{\rm PB} = \pm K_\pm\, ,\\
    &\big \{ J_i, K_j \big\}_{\rm PB}=0\,.
\end{aligned}
\end{equation}

\section{Spectrum of Minimally Coupled Kinetic Operators in AdS$_2\times \mathbf{S}^2$} \label{app:spectrumA2S2}

\noindent In this section, we briefly summarize the structure of the spectrum of the following differential operators
\begin{equation}\label{eq:opsS2&AdS2}
    \mathcal{D}^2 = -(\nabla - i A)^2 \,, \qquad \qquad \slashed{D}^2= - (\slashed{\nabla}-i \slashed{A})^2 \,,
\end{equation}
defined in both $\mathbf{S}^2$ and $\text{AdS}_2$ backgrounds. In particular, $\mathcal{D}^2 + m^2$ gives the kinetic operator for a spin-$0$ particle minimally coupled to gravity and a $U(1)$ gauge field, whereas $\slashed{D}^2 + m^2$ corresponds to (the square of the) kinetic operator for a similar spin-$\frac{1}{2}$ field. 

In our conventions, the $\mathbf{S}^2$ background is characterized by
\begin{align}\label{eq:MetricMaxwellSphere}
     ds^2 = R^2_{\mathbf{S}} \left(d\theta^2 + \sin^2\theta\, d\phi^2\right) \,,\qquad F = B\, \omega_{\mathbf{S}^2} = g\, \sin \theta\,d\theta \wedge d\phi\,, 
\end{align}
where $R_{\mathbf{S}}$ denotes the radius of the sphere, and $B = g/R_{\mathbf{S}}^2$ is the constant magnetic field, with $g \in \mathbb{Z}/2$ the quantized magnetic charge. The main strategy one adopts to determine the spectrum on $\mathbf{S}^2$ is observing that the Hamiltonian operator $H_{\mathbf{S}^2}$ of a spin-0 particle and the Dirac squared operator $\slashed{D}^2_{\mathbf{S}^2}$ are controlled by the Casimir of the sphere isometry group 
\begin{equation} \label{eq:S2operators}
    H_{\mathbf{S}^2} = \frac{1}{R_{\mathbf{S}}^2}\, \left(C_{SU(2)} - g^2\right) \,, \qquad  C_{SU(2)} = R^2\slashed{D}^2_{\mathbf{S}^2}+g^2-\frac14 \,.
\end{equation}
The eigenstates of \eqref{eq:S2operators} are accordingly organized in the standard discrete representations of SU(2). The heat kernel trace associated with $\mathcal{D}^2$, $\slashed{D}^2$, on the sphere can be computed using the eigenvalues $E_n$ and the degeneracies $d_n$
\begin{subequations}
\begin{align}
    & \underline{\text{spin-}0}: \qquad  E_n =  \frac{2 g}{R_{\mathbf{S}}^2} \left[ n + \frac{1}{2} + \frac{n(n+1)}{2g} \right]\,, \qquad  d_n = 2\left( g + n + \frac{1}{2} \right)\,, \label{summary:bosonS2}\\[2mm]
    & \underline{\text{spin-}1/2}: \qquad  E_n^2 =  \frac{1}{R^2_{\mathbf{S}}} n (n+2g)  \,, \qquad d_n =  4 n + (4 - 2  \delta_{n,0})\, g\,,\label{summary:fermionS2}
\end{align}
\end{subequations}
with  $n \in \mathbb{Z}_{\ge0}$, and it reads 
\begin{subequations}\label{eq:heatkernelS2_scalar&fermion}
\begin{align}
    & \mathcal{K}^{(0)}_{\mathbf{S}^2}(\tau)  = \sum_{n} d_n\, e^{-\tau E_n} = \sum_{n \geq 0} \left( 2n + 2g + 1 \right)\, e^{ -\frac{\tau}{R^2} \left( g + n(n+1+2g)\right)}\,,\\[2mm]
 & \mathcal{K}^{(1/2)}_{\mathbf{S}^2}(\tau)= \sum_{n} d_n\, e^{-\tau E^2_n} = 4\sum_{n} \left( n + g\right) e^{ -\frac{\tau}{R^2} n\left( n + 2g \right)} + 2g\,.
\end{align}
\end{subequations}

Similarly, the $\text{AdS}_2$ background is described by 
\begin{align}\label{eq:MetricMaxwellAdS}
ds^2 = \frac{R_{\rm{A}}^2}{\rho^2} \left(-dt^2 + d\rho^2\right)\,, \qquad F =  E\, \omega_{\text{AdS}_2}= e\, \frac{1}{\rho^2} dt \wedge d\rho\, ,
\end{align}
where $R_{\rm{A}}$ is the AdS radius, and $E = e/R_{\rm{A}}^2$ the constant electric field. In order to determine the AdS$_2$ spectrum we first consider the Euclidean version of the problem, i.e., an hyperbolic plane $\mathbb{H}^2$ threaded by a constant magnetic field. One obtains that the operators are controlled by the quadratic Casimir of SU(1,1):
\begin{equation} \label{eq:H2operators}
    H_{\mathbf{H}^2} =\frac{B}{2g}\, \left(C_{SU(1,1)} + g^2\right) \,, \qquad    C_{SU(1,1)} = R^2\slashed{D}^2_{\mathbb{H}^2}-g^2+\frac14\,, 
\end{equation}
The eigenstates of \eqref{eq:H2operators} are also organized in representations of  SU(1,1) but, being the group non-compact, we have now both a discrete set of states and the so-called continuous principal series. With such spectrum one can build the heat kernel trace of $\mathbb{H}^2$ which is then analytically continued to AdS$_2$ via a Wick rotation and the identification $g = -i e$. The heat kernel traces for the operators in \eqref{eq:opsS2&AdS2} can be thus computed via the eigenvalues $E_\lambda$ and energy densities\footnote{Notice that the fermionic density exhibits a simple pole at $\lambda=e$ and the spectral density, which must be positive definite, becomes strictly negative for $\lambda < e$. Taking the principal value and interpreting the integrand as a distribution is therefore crucial to resolves the negative density issue \cite{Castellano:2026ojb}.} $\rho_\lambda$
\begin{subequations}
\begin{align}
    & \underline{\text{spin-}0}: \qquad  E_\lambda =  \frac{1}{R_{\rm{A}}^2} \left( \lambda^2 + \frac{1}{4} - e^2 \right)\,, \qquad  \rho_B  = \frac{V_{\text{AdS}}}{2\pi R_{\rm{A}}^2} \frac{\lambda\,\sinh(2\pi \lambda)}{\cosh(2\pi \lambda) + \cosh(2\pi e)}\,,\label{summary:bosonAdS2} \\[2mm]
    & \underline{\text{spin-}1/2}: \qquad  E_\lambda^2 =  \frac{1}{R_{\rm{A}}^2} (\lambda^2 - e^2) \,, \qquad \rho_F = \text{P.V.}\, \bigg\{\frac{V_{\text{AdS}}}{\pi R_{\rm{A}}^2}  \frac{\lambda\,\sinh(2\pi \lambda)}{\cosh\left(2\pi \lambda\right) - \cosh\left(2\pi e\right)} \bigg\}\,,\label{summary:fermionAdS2}
\end{align}
\end{subequations}
with $\lambda \in \mathbb{R}_{>0}$, as follows
\begin{equation}\label{eq:heatkernelAdS2_scalar&fermion}
\mathcal{K}^{(0)}_{\rm AdS}(\tau) = \int_0^{\infty} 
d\lambda\, \rho_B(\lambda)\, e^{-\tau E_\lambda}\, ,\qquad \mathcal{K}^{(1/2)}_{\rm AdS}(\tau) = \int_0^{\infty} d\lambda\, \rho_F(\lambda)\ e^{-\tau E_\lambda^2} \,.
\end{equation}
%

%%%%%%%%%%%%%%
\bibliography{ref.bib}
\bibliographystyle{JHEP}

\end{document}